\documentclass[review,3p,times]{elsarticle}

\usepackage{amssymb,amsmath}
\usepackage{float,graphicx,lipsum}
\usepackage[percent]{overpic}
\usepackage{siunitx}
\DeclareSIUnit\USD{USD}
\DeclareSIUnit\kWh{kWh}
\DeclareSIUnit\bar{bar}
\DeclareSIUnit\milligram{mg}
\usepackage{soul}
\usepackage{url}
\usepackage[skip=4pt]{caption} 
\usepackage{multirow}
\usepackage{booktabs}
\usepackage{pifont}
\usepackage{hyperref}
\biboptions{sort&compress}
\usepackage{makecell}
\usepackage{array}
\newcolumntype{?}{!{\vrule width 1pt}}

\usepackage[dvipsnames]{xcolor}

\usepackage{enumitem}
\newlist{todolist}{itemize}{2}
\setlist[todolist]{label=$\square$}
\usepackage{pifont}

\usepackage{graphicx}
\usepackage{subcaption}
\usepackage{lineno}

\journal{}

\begin{document}

\begin{frontmatter}

\title{
High-Throughput Bayesian Optimization of Cement-Salt Hydrates Composites for Seasonal Thermochemical Energy Storage
}

\author[Denerg]{Alessio Mondello}

\author[Denerg]{Giulio Barletta}

\author[Disat]{Luca Lavagna}

\author[Denerg]{Matteo Fasano}

\author[Disat]{Matteo Pavese}

\author[Denerg,INRiM]{Eliodoro Chiavazzo\corref{cor}}

\affiliation[Denerg]{
            organization={Department of Energy, Politecnico di Torino},%Department and Organization
            addressline={Corso Duca degli Abruzzi, 24}, 
            city={Torino},
            postcode={10129}, 
            country={Italy}}

\affiliation[Disat]{
            organization={Department of Applied Science and Technology, Politecnico di Torino},%Department and Organization
            addressline={Corso Duca degli Abruzzi, 24}, 
            city={Torino},
            postcode={10129}, 
            country={Italy}}

\affiliation[INRiM]{
            organization={Istituto Nazionale di Ricerca Metrologica},%Department and Organization
            addressline={Strada delle Cacce, 91}, 
            city={Torino},
            postcode={10135}, 
            country={Italy}}

\cortext[cor]{\href{mailto:eliodoro.chiavazzo@polito.it}{eliodoro.chiavazzo@polito.it}}
% \fntext[samecont]{Authors contributed equally}

\begin{abstract}
Thermochemical energy storage (TCES) based on salt hydrates is a promising route for seasonal heat storage; however, the design of practical sorbent materials remains challenging due to a non-trivial coupling between composition, synthesis feasibility, performance, and cost.
Here, focusing on salt-into-matrix cement-based composites, we demonstrate that a high-throughput experimental framework based on Bayesian optimization (BO) can be used to orchestrate the optimization process of composite materials for low-temperature TCES.
The explored design space is defined by salt type, salt concentration, water-to-cement ratio, and additive-to-cement ratio, while two competing objectives are pursued in parallel, namely the specific energy and the specific energy cost.
The BO-guided campaign identified Pareto-optimal composites based on CaCl$_2$, Zn(NO$_3$)$_2$, and LiCl, highlighting the promise of cement-salt combinations that have been only marginally explored, or not previously reported, in cement-based TCES systems.
The best-performing formulation (LiCl-based), achieved an average specific energy of about $\SI{458}{\kilo\joule\per\kilo\gram}$, whereas CaCl$_2$- and Zn(NO$_3$)$_2$-based composites showed lower but still competitive specific energy values combined with more favorable specific energy cost.
Overall, the optimized formulations improved the specific energy of previously developed cement-based materials by up to a factor of five, although it remains below that of state-of-the-art composites based on silica gel and expanded vermiculite.
Nonetheless, the present materials, notably CaCl$_2$- and Zn(NO$_3$)$_2$-based composites, offer an attractive cost-to-performance balance, highlighting BO as an effective strategy for accelerated TCES materials discovery.
\end{abstract}


\end{frontmatter}
% \linenumbers

%%%%%%%%%%%%%%%%%%%%% main text

%%%%%%%%%%%%%%%%%%%%%%%%%%%%%%%%%%%%%%%%%%%%%%%%%%%
\section{Introduction}\label{sec:introduction}
Thermal energy storage (TES) is widely regarded as a key enabling technology for the large-scale integration of intermittent renewable energy sources towards decarbonization of the energy sector both in the industrial and civil sectors~\cite{skevi_reviewing_2025,spietz_thermochemical_2025,whitepaperIWG52024}.
A major challenge lies in the temporal mismatch between energy availability and demand.
In particular, solar thermal energy exhibits both a daily peak during daylight hours and a seasonal peak in summer, whereas heat demand is typically higher in the evening and during winter.
While short- and medium-term thermal energy storage technologies are characterized by a higher technological maturity~\cite{morciano2025trending,neri2020numerical}, a strong research effort is still needed to develop affordable long-term TES systems capable of storing thermal energy on a seasonal timescale~\cite{skevi_reviewing_2025,spietz_thermochemical_2025}.

More specifically, three main families are typically identified according to the main mechanism by which heat is stored and released.
Sensible heat storage (SHS) technologies rely on temperature variations of a storage mean;
latent heat storage (LHS) technologies exploit phase transitions;
and thermochemical energy storage (TCES) technologies store heat through reversible chemical or physical interactions~\cite{skevi_reviewing_2025, spietz_thermochemical_2025}.
SHS and LHS are well-established for short- and medium-term applications; however, they suffer from thermal losses and relatively low energy densities, which limit their suitability for long-term storage~\cite{clark_experimental_2021, spietz_thermochemical_2025}.
TCES systems, although not yet fully mature, are an attractive option for long-term applications due to the intrinsic advantages of their storage mechanism~\cite{skevi_reviewing_2025, spietz_thermochemical_2025}.
TCES exhibit higher volumetric energy densities than SHS and LHS, enabling a reduction in the required plant footprint~\cite{spietz_thermochemical_2025}.
Moreover, long-term storage is feasible because energy is stored in the form of chemical potential: if the physical separation between adsorbate and adsorbent is properly managed, the stored energy remains preserved and is released only when the two active components are brought into contact~\cite{skevi_reviewing_2025, clark_experimental_2021}.

For low-temperature TCES, water vapor is the most commonly used adsorbate due to its safety, availability, and favorable operating conditions~\cite{zhang_research_2025}.
The selection of the most suitable sorbent material remains an open and highly active research topic within the scientific community~\cite{spietz_thermochemical_2025, liu_systematic_2025}.
The optimal choice depends strongly on the type of TCES system (open or closed) and on the operating conditions, such as available charging temperature and discharging humidity~\cite{yang_salt_2023, liu_systematic_2025}.
Hydrated salts are among the most extensively investigated sorbent materials because of their high theoretical energy density~\cite{spietz_thermochemical_2025, yang_salt_2023, zhang_research_2025}.
These salts store and release heat through successive dehydration and hydration cycles; however, their practical implementation is hindered by well-known drawbacks, including deliquescence, swelling, mechanical instability, and limited cyclability~\cite{clark_experimental_2021, yang_salt_2023, liu_systematic_2025}.
To overcome these limitations, the incorporation of hydrated salts into a stable porous host matrix has been widely explored in the literature, and the resulting materials are commonly referred to as ``composite salt in porous matrix'' (CSPM) systems~\cite{gordeeva_composites_2012}.
The porous matrix provides mechanical support, promotes salt dispersion, mitigates agglomeration, and could improve heat and mass transfer within the material~\cite{zhang_research_2025, liu_systematic_2025}.
In addition to conventional porous matrices such as silica gel, zeolites, vermiculite, and other mineral supports~\cite{zhang_research_2025, liu_systematic_2025}, cement-based materials have recently been investigated as innovative and low-cost host matrices for hydrated salts~\cite{clark_experimental_2021, salustro_thermal_2024, mohapatra_salt_2023, skevi_reviewing_2025}.
The use of cement as host matrix offers several potential advantages, including robust structural integrity and the presence of a multiscale porous network capable of accommodating and dispersing salt~\cite{clark_experimental_2021, salustro_thermal_2024, lavagna_insight_2022}.

Despite this potential, the formulation of cement--salt composites for TCES remains a highly non-trivial design problem.
The performance of the final material depends on multiple coupled variables, including the salt chemistry, salt loading, binder composition, and additive content, which jointly affect thermochemical performance, processability, stability, and cost of the materials.
As a result, the associated search space is combinatorial and strongly non-linear.
A full experimental exploration of this space would require thousands of synthesis--characterization cycles, each involving sample preparation, curing, granulation, and sorption analysis over timescales of weeks.
This makes conventional trial-and-error strategies, one-factor-at-a-time studies, or even dense grid-based screening impractical for the efficient identification of optimal formulations.

In recent years, data-driven optimization strategies have emerged as powerful tools for accelerating materials discovery and reducing the experimental burden associated with expensive searches~\cite{ling_high-dimensional_2017,lookman_active_2019,gomez-bombarelli_automatic_2018,janet_accelerating_2018,jin_machine-learning-assisted_2022}.
Among these approaches, Bayesian optimization (BO) has proven particularly effective for navigating large experimental spaces under tight resource constraints.
By combining probabilistic surrogate models with acquisition functions that balance exploration and exploitation, BO enables the efficient identification of optimal materials or processing conditions using relatively small datasets~\cite{frazier_tutorial_2018,brochu_tutorial_2010,shahriari_taking_2016}.
This methodology has been successfully applied to a wide range of materials problems, enabling sample-efficient discovery of high-performance materials in chemistry, catalysis, and energy storage systems~\cite{guo_bayesian_2023,yang_bayesian_2022,khatamsaz_bayesian_2023,langner_beyond_2020,agarwal_discovery_2021,zhang_exploring_2023,bonke_multi-variable_2024,zhang_optimizing_2023,doan_quantum_2020,niu_accelerated_2025,yik_accelerating_2025,li_computational_2025}.
A few examples of using sequential learning algorithms for the optimization of thermal energy storage materials have been reported in purely numerical studies~\cite{trezza_minimal_2022}.
However, its application to TCES materials remains largely unexplored, especially for cement-based salt hydrate composites, where synthesis/stability constraints and multi-objective trade-offs play a central role.

In this work, building on the preliminary studies by Lavagna et al.~\cite{lavagna_cementitious_2020}, Clark et al.~\cite{clark_experimental_2021}, and Salustro et al.~\cite{salustro_thermal_2024}, we develop a high-throughput, data-driven experimental framework based on BO to guide the formulation of cement-based salt hydrate composites for seasonal TCES.
The method explores a four-dimensional design space defined by salt type, salt concentration, water-to-cement ratio, and additive-to-cement ratio, while simultaneously targeting high specific energy and low specific energy cost.
This approach enables the efficient identification of promising composite formulations while revealing synthesis boundaries and performance trade-offs within the combinatorial design space.
To the best of our knowledge, this is the first study to apply Bayesian optimization to the multi-objective design of cement--salt composites for TCES.
In addition to optimizing formulation variables, the present study explores a broader set of active salts than those most commonly considered in cement-based TCES composites~\cite{skevi_reviewing_2025,liu_systematic_2025, clark_experimental_2021, salustro_thermal_2024, lavagna_cementitious_2020}, including LiCl, CaCl$_2$, and Zn(NO$_3$)$_2$, thereby extending the accessible chemical design space of these materials.

This manuscript is organized in sections as follows.
Section~\ref{sec:methods} describes the experimental and computational methodology adopted in this work, including the BO framework, composite synthesis, adsorption measurements, and thermodynamic modeling.
Section~\ref{sec:results} presents and discusses the optimization outcomes, the adsorption and thermodynamic analyses of the selected composites, and their comparison with relevant literature benchmarks.
Finally, Section~\ref{sec:conclusions} summarizes the main conclusions and highlights possible future developments.

\section{Methods}\label{sec:methods}
\subsection{Materials}\label{ssec:materials}
In this work, all composites were synthesized using ordinary Portland cement (PC) ``Type I 52,5R'' supplied by Buzzi Unicem.
The hydrated salts employed as active sorbents included
magnesium sulfate heptahydrate (MgSO\textsubscript{4}$\cdot$7H\textsubscript{2}O),
copper sulfate pentahydrate (CuSO\textsubscript{4}$\cdot$5H\textsubscript{2}O),
aluminium sulfate octadecahydrate (Al\textsubscript{2}(SO\textsubscript{4})\textsubscript{3}$\cdot$18H\textsubscript{2}O),
aluminium potassium sulfate dodecahydrate (AlK(SO\textsubscript{4})\textsubscript{2}$\cdot$12H\textsubscript{2}O),
calcium chloride dihydrate (CaCl\textsubscript{2}$\cdot$2H\textsubscript{2}O),
magnesium chloride hexahydrate (MgCl\textsubscript{2}$\cdot$6H\textsubscript{2}O),
lithium chloride monohydrate (LiCl$\cdot$H\textsubscript{2}O),
potassium carbonate sesquihydrate (K\textsubscript{2}CO\textsubscript{3}$\cdot$1.5H\textsubscript{2}O),
magnesium nitrate hexahydrate (Mg(NO\textsubscript{3})\textsubscript{2}$\cdot$6H\textsubscript{2}O),
and zinc nitrate hexahydrate (Zn(NO\textsubscript{3})\textsubscript{2}$\cdot$6H\textsubscript{2}O),
all purchased from Merck.
Strontium bromide hexahydrate (SrBr\textsubscript{2}$\cdot$6H\textsubscript{2}O) was purchased from Thermo Fisher Scientific.
Mastermatrix UW 444 (Master Builder Solutions) was used as the anti-settling additive.
This commercial anti-washout powder was incorporated as a thickening agent to mitigate bleeding phenomena during paste preparation.
All saline solutions and cement pastes were prepared using freshly produced deionized water.

\subsection{Optimization Strategy}\label{ssec:optimization}
\begin{table}[t]
    \centering
    \caption{Design variables defining the four-dimensional formulation space explored in this work.
    The parameter $s/s_{max} \; \left[\SI{}{\kilogram\per\kilogram}\right]$ denotes the mass of salt dissolved in water relative to the maximum mass of salt soluble in \SI{100}{\gram} of water at \SI{20}{\degreeCelsius}~\cite{haynes_crc_2014}; $w/c \; \left[\SI{}{\kilogram\per\kilogram}\right]$ is the mass ratio between water and dry cement powder; $a/c \; \left[\SI{}{\gram\per\kilogram}\right]$ is the mass ratio of additive relative to dry cement powder.}
    \begin{tabular}{p{3cm}|lp{3.5cm}r}
         Variable & Variable type & List & Range \\
         \hline
         \multirow{11}{4em}{Salt type} & \multirow{11}{4em}{Categorical} & [Al\textsubscript{2}(SO\textsubscript{4})\textsubscript{3}$\cdot$18H\textsubscript{2}O & \multirow{11}{4em}{---} \\
         & & AlK(SO\textsubscript{4})\textsubscript{2}$\cdot$12H\textsubscript{2}O & \\
         & & CaCl\textsubscript{2}$\cdot$2H\textsubscript{2}O & \\
         & & CuSO\textsubscript{4}$\cdot$5H\textsubscript{2}O & \\
         & & K\textsubscript{2}CO\textsubscript{3}$\cdot$1.5H\textsubscript{2}O & \\
         & & LiCl$\cdot$H\textsubscript{2}O & \\
         & & Mg(NO\textsubscript{3})\textsubscript{2}$\cdot$6H\textsubscript{2}O & \\
         & & MgCl\textsubscript{2}$\cdot$6H\textsubscript{2}O & \\
         & & MgSO\textsubscript{4}$\cdot$7H\textsubscript{2}O & \\
         & & SrBr\textsubscript{2}$\cdot$6H\textsubscript{2}O & \\
         & & Zn(NO\textsubscript{3})\textsubscript{2}$\cdot$6H\textsubscript{2}O] & \\
         $s/s_{max}$ $\left[\SI{}{\kilogram\per\kilogram}\right]$ & Continuous & --- & 0.10 -- 0.90 \\
         $w/c$ $\left[\SI{}{\kilogram\per\kilogram}\right]$ & Continuous & --- & 0.70 -- 1.50 \\
         $a/c$ $\left[ \SI{}{\gram\per\kilogram} \right]$ & Continuous & --- & 0.0 -- 30.0
    \end{tabular}
    \label{tab:variables}
\end{table}

\begin{figure}[t]
    \centering
    \includegraphics[width=\linewidth]{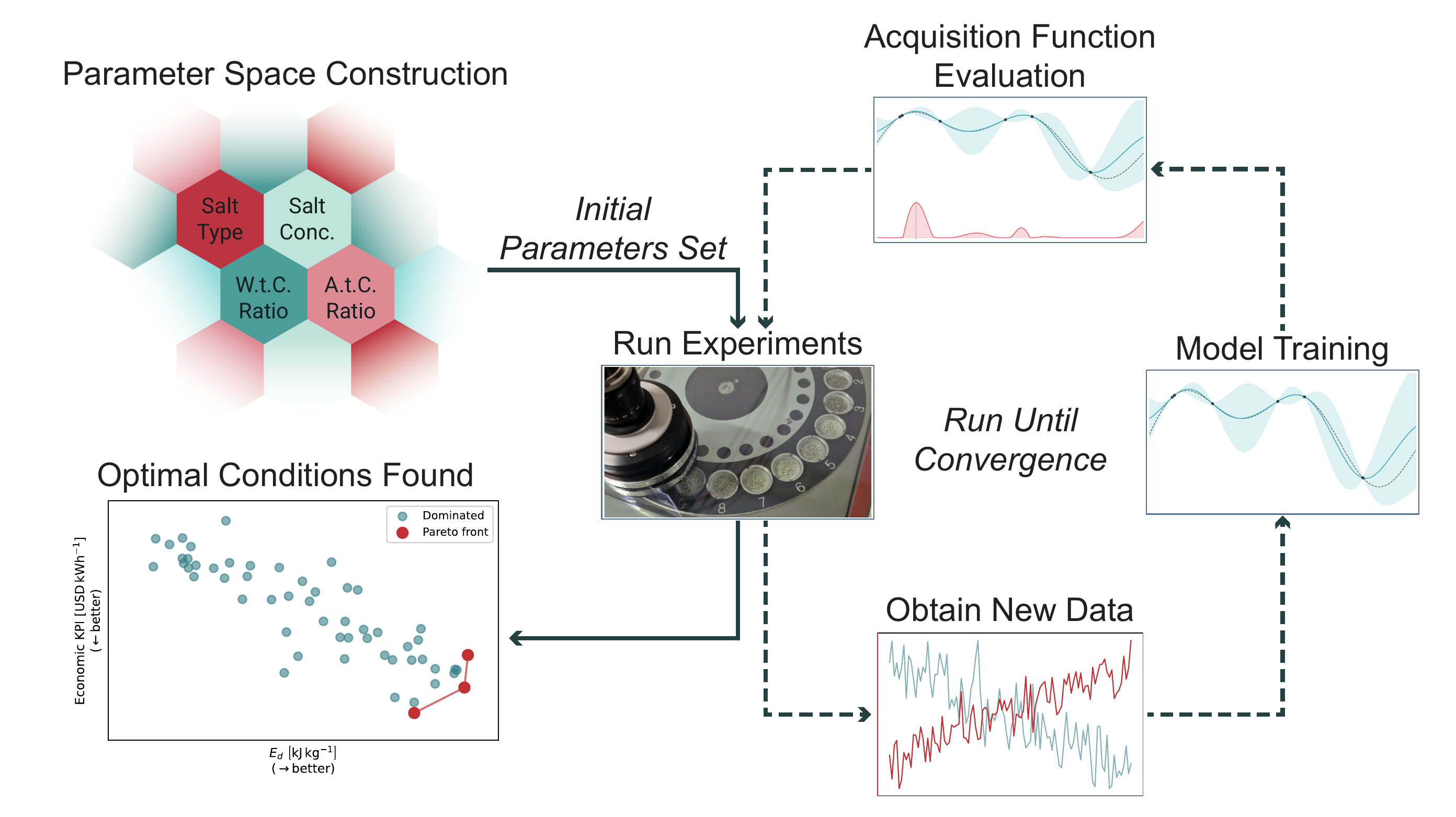}
    \caption{Schematic representation of the BO framework employed in this work.
    The four-dimensional parameter space is defined by the salt type, salt concentration $\left( s/s_{max} \; \left[\SI{}{\kilogram\per\kilogram} \right)\right]$), water-to-cement ratio $\left( w/c \; \left[\SI{}{\kilogram\per\kilogram}\right] \right)$, and additive-to-cement ratio $\left( a/c \left[ \SI{}{\gram\per\kilogram} \right] \right)$.
    The BO loop begins with an initial set of experimental points used to train a GP surrogate model.
    The model is iteratively updated as new data are collected, with acquisition functions guiding the selection of subsequent experiments to balance exploration and exploitation.
    Although the loop can in principle be iterated until a satisfactory optimization outcome is achieved, progressively refining the prediction of promising conditions, in practice the campaign may also be stopped according to experimental or project-specific constraints.
    }
    \label{fig:optimization}
\end{figure}

The optimization of the TCES composites was carried out through an adaptive BO workflow~\cite{brochu_tutorial_2010,frazier_tutorial_2018,shahriari_taking_2016}, designed to efficiently explore a multidimensional formulation space under multiple competing objectives.
The design space was defined by four experimental variables, namely the salt type (categorical descriptor corresponding to the hydrated salt used), the salt concentration $\left( s/s_{max} \; \left[\SI{}{\kilogram\per\kilogram} \right)\right]$), the water-to-cement ratio $\left( w/c \; \left[\SI{}{\kilogram\per\kilogram}\right] \right)$, and the additive-to-cement percentage ratio $\left( a/c \left[ \SI{}{\gram\per\kilogram} \right] \right)$.
The salt hydrate candidates chosen in this work were selected from literature~\cite{lin_applications_2021, van_essen_characterization_2009, zbair_survey_2021, richter_systematic_2018, trausel_review_2014, donkers_review_2017, ntsoukpoe_systematic_2014, clark_experimental_2022} based on their theoretical specific energy $E_d$, cyclability, regeneration temperature, and non-toxicity.
The resulting set was intentionally designed to cover a chemically diverse range of TCES-active salts, including both MgSO$_4$, already considered in cement-salt composite sorbents~\cite{lavagna_cementitious_2020, salustro_thermal_2024} and salts that, to the best of our knowledge, had not yet been investigated as active phases in cement-based TCES composites, such as LiCl, CaCl$_2$, and Zn(NO$_3$)$_2$~\cite{skevi_reviewing_2025,liu_systematic_2025,spietz_thermochemical_2025, clark_experimental_2021}.
This choice was aimed not only at exploring promising but under-investigated regions of the formulation space, but also at testing whether cement-based matrices could effectively host salts with potentially more favorable thermochemical behavior than those most commonly considered so far.
The full list of salt types, along with the investigated ranges of the continuous variables, is reported in Table~\ref{tab:variables}.

These descriptors define a four-dimensional formulation space that captures the compositional variability governing the thermochemical performance, processability, and cost of the resulting composites.
The optimization targeted two conflicting objectives:
(i) maximization of the specific energy $E_d \; \left[\SI{}{\kilo\joule\per\kilogram}\right]$ of the composite, and
(ii) minimization of an economic key performance indicator (KPI) $\left[\SI{}{\USD\per\kWh}\right]$, defined as the specific energy cost of the composite.
Rather than using a single multi-objective BO algorithm, these objectives were handled through two parallel BO pipelines, each optimized independently.
At every round, 10 candidates were proposed by the $E_d$-oriented pipeline and 10 by the KPI-oriented pipeline.
The full set of experimentally evaluated composites was then analyzed \textit{a posteriori} through Pareto-front construction.

Starting from a preliminary exploratory dataset of 22 manually selected composites, each optimization round comprised 20 experiments, corresponding to 20 unique formulations selected according to a multi-objective BO strategy.
To accelerate the exploration of the formulation space, redundant or experimentally indistinguishable compositions were systematically excluded.
For each objective, the surrogate models were implemented as Gaussian Processes (GPs)~\cite{brochu_tutorial_2010,seeger_gaussian_2004} trained on all the experimental data available up to that iteration.
To increase model diversity and reduce kernel bias, two covariance functions were used in parallel, namely the radial basis function (RBF) kernel and the Matérn kernel.
The corresponding hyperparameters, in particular the characteristic length scales, were re-optimized at every iteration and constrained within the range $\left[ 10^{-8}, 10 \right]$ in normalized units to maintain numerical stability and prevent over- or under-fitting of local features.
Within each kernel, five acquisition strategies were considered: Probability of Improvement (PI)~\cite{kushner_new_1964,torn_global_1989,jones_taxonomy_2001}, Expected Improvement (EI)~\cite{mockus_application_1994,jones_efficient_1998}, and Lower Confidence Bound (LCB)~\cite{cox_statistical_1992} with three values of the exploration parameter $\left( \kappa=\left[ 1.0, 4.0, 10.0 \right] \right)$.
This ensemble of different kernels and acquisition functions fostered a balance between exploration of uncertain regions and exploitation of high-performing areas of the design space.
A complete mapping of the 20 BO trial branches considered at each optimization round, including objective, kernel, and acquisition-function assignments, is reported in the Supplementary Information.

To avoid near-duplicate suggestions, which can arise especially from more exploitative acquisition criteria, a minimum-spacing filter was applied to the Euclidean (2-norm) distance between candidate points in the normalized continuous input space.
The distance threshold was set to 0.2, a value empirically chosen to reflect the experimental resolution of the formulation process and to maintain sufficient diversity among proposed candidates.
This constraint ensured that each batch explored genuinely distinct regions of the design space, thereby maximizing information gain from each experimental cycle~\cite{martin_autonomous_2025}.

At the end of each optimization round, the surrogate models were retrained using the updated cumulative dataset, and the cycle was repeated to progressively refine the representation of the underlying response surfaces.
A schematic overview of the BO workflow, including model training, acquisition, candidate selection, and experimental validation, is provided in Figure~\ref{fig:optimization}.

All computational work was carried out in Python (v3.10.15) using the scikit-optimize~\cite{head_scikit-optimizescikit-optimize_2021} and scikit-learn~\cite{pedregosa_scikit-learn_2011} libraries for BO, in combination with standard packages such as NumPy~\cite{harris_array_2020}, pandas~\cite{mckinney_data_2010}, SciPy~\cite{virtanen_scipy_2020}, and matplotlib~\cite{hunter_matplotlib_2007} for data handling and visualization.
This framework ensured both flexibility in defining custom optimization routines and reproducibility of the entire workflow.

\subsection{Synthesis}\label{ssec:synthesis}
All composites were synthesized according to a standard protocol adapted from the in-situ preparation method proposed by Lavagna et al.~\cite{lavagna_cementitious_2020}.
First, the required amounts of salt and water were weighed, and the salt was dissolved in water inside a sealed Falcon tube.
The mixture was shaken until a clear saline solution was obtained and no undissolved crystals remained.
In parallel, the required amounts of cement powder and anti-settling additive were weighed, combined in a separate Falcon tube, and homogenized in a Turbula mixer for at least \SI{30}{\minute}.
The dry mixture was then transferred to a polyethylene beaker and mixed with the saline solution using a vertical mixer equipped with a four-blade impeller for a minimum of \SI{60}{\second}, so as to ensure uniform dispersion.
The mixing speed was kept constant across all samples, except for formulations exhibiting rapid setting, for which a more vigorous mixing was applied to achieve adequate homogenization.
The resulting cement paste was poured into silicone molds $\left(\SI{35}{} \times \SI{35}{} \times \SI{35}{\milli\meter}\right)$.
For formulations prone to rapid thickening, the paste was transferred using a spatula and manually leveled to completely fill the mold.
The molded samples were then cured in a glove box at 100\% relative humidity (RH) and ambient temperature for 7 days, followed by overnight pre-drying in a Binder FED laboratory oven at \SI{90}{\degreeCelsius}.
The pre-dried bulk specimens were mechanically reduced to granules with a hammer and subsequently sieved through \SI{1}{} and \SI{2}{\milli\meter} meshes.
Only granules with sizes within this size range were retained for further characterization.
This protocol was designed to ensure sufficient reproducibility and throughput for batch-wise screening within the BO campaign.

Considering the nature of the composites investigated in this work, the salt content could not be evaluated using the conventional procedure commonly adopted for impregnated salt-in-matrix composites \cite{grekova_composite_2017, courbon_further_2017, brancato_experimental_2021}.
In that approach, the dry matrix is weighed before impregnation and then the dry composite is weighed after impregnation, so that the salt content is determined from the mass increase associated with salt loading.
In the present case, however, the salt is introduced directly during the cement-based composite synthesis, and a pre-formed dry matrix is not available.
Therefore, a different approach was adopted, in which the salt content was calculated from the initial dry powder masses and the amount of water stoichiometrically consumed during Portland cement hydration.

Assuming that all the salt we incorporate remains embedded in the final matrix, neglecting the small losses that can be associated with the bleeding phenomena, the salt content was calculated as:
\begin{equation}
    \mathrm{Salt\ content} = 
    \frac{m_\mathrm{salt,anhydrous}}
    {m_\mathrm{PC} + m_\mathrm{water,reacted} + m_\mathrm{additive} + m_\mathrm{salt,anhydrous}}
\label{eq:saltcontent}
\end{equation}
where $m_{salt,anhydrous}$ is the theoretical mass of anhydrous salt in the composite, calculated from the actual mass of hydrated salt introduced into the mixture; $m_{PC}$ is the weighed mass of PC powder used in the mixture; $m_{additive}$ is the weighed mass of additive used in the mixture; and $m_{water,reacted}$ is the mass of the consumed stoichiometric water during cement hydration, assumed to be \SI{23}{\gram} of water per \SI{100}{\gram} of cement powder~\cite{lavagna_insight_2022, lea_leas_2004}. 

\subsection{Adsorption Isotherms}\label{ssec:adsorption}
The adsorption behavior of the composites, used to derive their corresponding $E_d$, was evaluated through water adsorption isotherms measured using a ``SPSx-1$\mu$ High Load'' dynamic vapor sorption (DVS) analyzer (ProUmid).
Prior to analysis, the granulated composites were dried at \SI{90}{\degreeCelsius} for \SI{24}{\hour} in a laboratory oven, then transferred into \SI{1.2}{\milli\liter} microcentrifuge tubes and sealed to prevent exposure to ambient humidity.
For each measurement, the granules were poured into a pre-tared aluminum pan (\SI{33}{\milli\meter} diameter) in an amount sufficient to cover approximately one third of the pan surface. 
The mass recorded under these conditions was taken as the reference weight at 0\% RH.
The isotherms were collected at \SI{20}{\degreeCelsius}.
Each DVS run allowed the simultaneous analysis of 20 composites, enabling high-throughput screening of the formulation space.
For the main dataset, RH was varied from 5\% to 50\% RH in 5\% increments, yielding 10 equilibrium points per isotherm.
The weighing interval was set such that the instrument recorded one mass measurement every \SI{14}{\minute}.
Equilibrium was defined as a mass variation rate below 0.1\% over \SI{40}{\minute}, a threshold selected to compromise between measurement accuracy and experimental throughput.
For the initial set of 22 composites used to seed the BO campaign, isotherms were instead collected over a wider humidity range, from 10\% to 80\% RH in 10\% increments.

\subsection{Physical and Thermodynamic Modeling}\label{ssec:modelling}
\begin{figure}[t]
    \centering
    \includegraphics[width=0.70\linewidth]{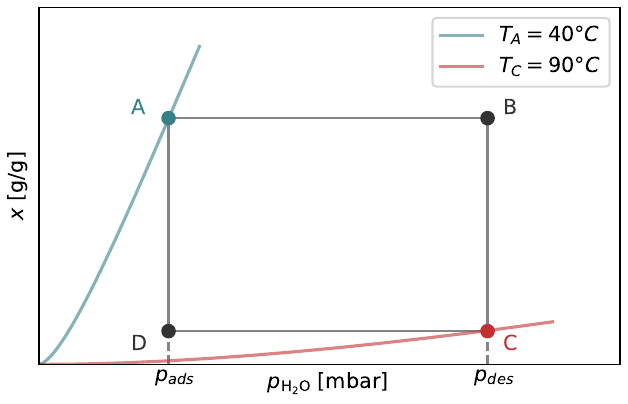}
    \caption{Schematic representation of the thermodynamic cycle adopted to estimate the specific energy of the composites. 
    The two solid curves represent the Dubinin--Astakhov based adsorption isotherms at the ambient operating temperature $T_A=\SI{40}{\degreeCelsius}$ and at the charging temperature $T_C=\SI{90}{\degreeCelsius}$.
    Point A corresponds to adsorption at $T_A$ and $p_{ads}$, whereas point C corresponds to desorption at $T_C$ and $p_{des}$.
    The cycle uptake variation is defined as $\Delta x_{\mathrm{cycle}}=x_A-x_C$.
    Points B and D are identified as the intersections of the isosteres $x_A$ and $x_C$ with the desorption pressure $p_{\mathrm{des}}$ and the adsorption pressure $p_{\mathrm{ads}}$, respectively.}
    \label{fig:cycle_scheme}
\end{figure}
To quantify the specific energy of the composites, a combined physical and thermodynamic modeling approach was applied.
In this work, the $E_d$ is identified with the cycled heat $Q_u$ expressed per unit mass of dry sorbent $\left[\SI{}{\kilo\joule\per\kilo\gram\of{dry\;sorbent}}\right]$.

All pressure and temperature data were first converted into consistent units (\SI{}{\milli\bar} and \SI{}{\kelvin}).
Assuming that the Polanyi adsorption potential theory can be applied, the experimental adsorption isotherms were fitted using the mass-based Dubinin–Astakhov (DA) equation~\cite{aristov_applying_2023}:
\begin{equation}
    x = x_0 \, \mathrm{exp} \left[ - \left( \frac{A}{E} \right)^n \right]
\label{eq:d-a}
\end{equation}
where $x$ is the equilibrium water uptake, defined as mass of adsorbed water per unit mass of dry sorbent $\left[\SI{}{\kilogram\of{water}\per\kilogram\of{dry\;sorbent}}\right]$, $x_0$ is the maximum uptake capacity of the material, $A$ is the adsorption potential, $E$ is the characteristic adsorption energy of the system, and $n$ is an empirical parameter.
Parameters $x_0$, $E$, and $n$ were obtained by fitting, while $A$ was computed at each pressure according to the Polanyi relation~\cite{aristov_applying_2023}:
\begin{equation}
    A = -RT \, \mathrm{ln}(p/p_0)
\label{eq:polanyi}
\end{equation}
where $R$ is the gas constant, $p$ is the equilibrium vapor pressure of the adsorbate, and $p_0$ is the corresponding saturation pressure, taken from literature sources~\cite{haynes_crc_2014}.

To evaluate the agreement between the modeled and experimental isotherms, the coefficient of determination, $R^2$, was calculated for each curve as:
\begin{equation}
    R^2 = 1 - \frac{\sum_i \left(x_{i,\mathrm{exp}} - x_{i,\mathrm{mod}}\right)^2}
    {\sum_i \left(x_{i,\mathrm{exp}} - \overline{x}_{\mathrm{exp}}\right)^2}
\label{eq:r2}
\end{equation}
where $x_{i,\mathrm{exp}}$ and $x_{i,\mathrm{mod}}$ are the experimental and modeled water uptake values at the same relative humidity point, respectively, and $\overline{x}_{\mathrm{exp}}$ is the mean experimental uptake.

After fitting the experimental isotherms at \SI{20}{\degreeCelsius}, the DA model was used to generate isotherms at higher temperatures by applying Polanyi's potential approach.
In this procedure, the adsorption potential $A$ is kept constant, while the corresponding equilibrium pressure is recalculated at the target temperatures.
Two temperatures were considered: \SI{40}{\degreeCelsius}, representing the operating temperature $T_A$, and \SI{90}{\degreeCelsius}, corresponding to the charging temperature $T_C$.
This procedure yielded DA-based isotherms at both $T_A$ and $T_C$.
A schematic representation of the resulting thermodynamic cycle is shown in Figure~\ref{fig:cycle_scheme}.

Two characteristic operating points, denoted as A and C, were then identified on the modeled isotherms.
Point A was defined on the \SI{40}{\degreeCelsius} isotherm at a pressure ($p_{ads}$) equal to the saturation vapor pressure at \SI{10}{\degreeCelsius}, namely \SI{12.3}{\milli\bar}, taken as a representative winter ambient temperature.
Point C was evaluated on the \SI{90}{\degreeCelsius} isotherm at a pressure ($p_{des}$) equal to the saturation vapor pressure at \SI{30}{\degreeCelsius}, namely \SI{42.5}{\milli\bar}, taken as a representative summer ambient temperature.
The corresponding water uptakes, $x_A$ and $x_C$, were obtained by evaluating Equation~\ref{eq:d-a} at these pressures.

To estimate the isosteric heat of adsorption $q_\mathrm{st}$, five uniformly spaced water uptake levels between points A and C were selected.
For each uptake value, the equilibrium pressures at $T_A$ and $T_C$ were computed by numerically solving Equation~\ref{eq:d-a}.
The isosteric heat at each uptake was then calculated using the Clausius-Clapeyron equation:
\begin{equation}
    q_\mathrm{st} = R \, \frac{T_1T_2}{T_2 - T_1} \mathrm{ln}{ \left( \frac{p_2}{p_1} \right)},
\label{eq:c-c}
\end{equation}
where $T_1=T_A$, $T_2=T_C$, and $p_1$, $p_2$ are the corresponding equilibrium pressures predicted by the DA model.
The final value of $\overline {q_\mathrm{st}}$ was taken as the average over the five selected uptake levels.
A step-by-step derivation of the thermodynamic cycle modeling procedure, including the combination of the DA equation with the Polanyi relation, the Clausius--Clapeyron derivation, and the identification of the operating points and cycled heat, is provided in the Supplementary Information.

The specific energy $E_d$ of the composites is taken to be the cycled heat $Q_u$, calculated as:
\begin{equation}
    E_d \equiv Q_u = \overline {q_{st}} \, \frac{\Delta x_{\mathrm{cycle}}}{M_\mathrm{H_2O}}
\label{eq:cycled_heat}
\end{equation}
where $\Delta x_{\mathrm{cycle}}=x_A-x_C$ is the difference in water uptake between points A and C, and $M_\mathrm{H_2O}$ is the molar mass of water.
In addition to $E_d$, the specific energy cost was adopted as an economic KPI to quantify the cost-effectiveness of each composite formulation.
The KPI $\left[\SI{}{\USD\per\kWh}\right]$ was computed as a function of the material cost per unit mass and normalized by the corresponding $E_d$:
\begin{equation}
    \mathrm{KPI}=\frac{C_{\mathrm{mat}}}{E_d},
    \label{eq:kpi}
\end{equation}
where $C_\mathrm{mat}$ represents the estimated material cost of the composite, accounting for the contributions of cement, salt hydrate, water, and additives.
\section{Results and Discussion}\label{sec:results}
\subsection{Optimization in the Combinatorial Space}\label{ssec:screening}
\begin{figure}[h!]
    \centering
    \includegraphics[width=\linewidth]{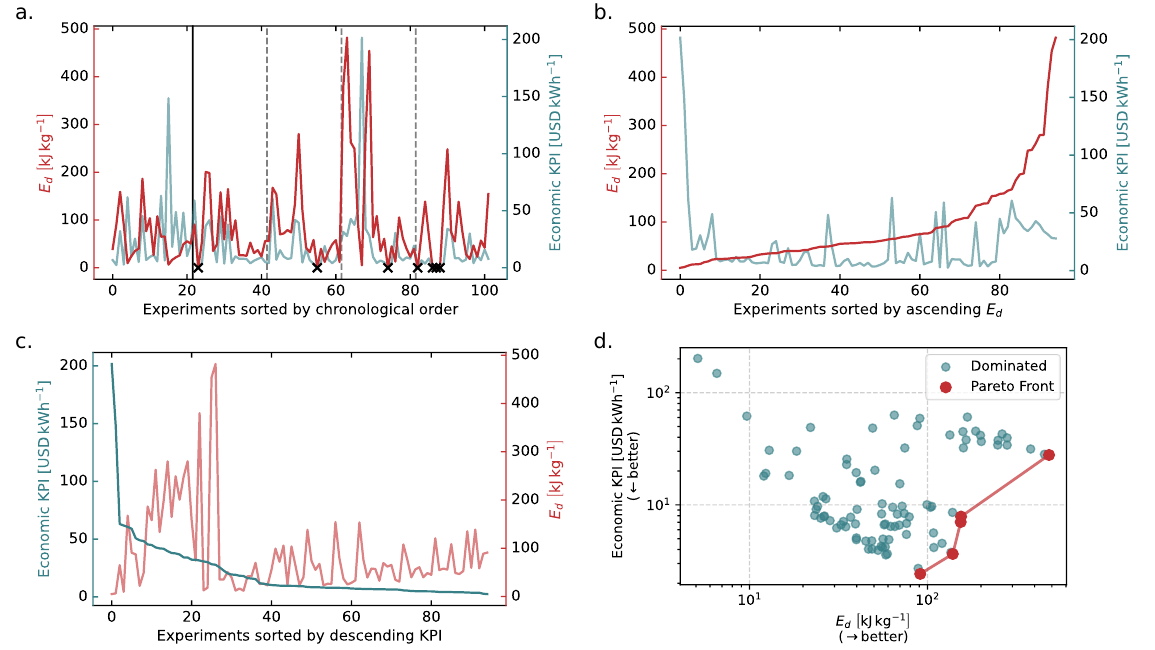}
    \caption{
    Summary of the BO results across all synthesized cement--salt composites.
    (a) $E_d$ (red) and economic KPI (blue) for all experiments, shown in chronological order.
    A solid black vertical line marks the boundary between the initial set of arbitrarily selected formulations and the beginning of the BO-guided campaign, while dashed gray vertical lines appear every 20 experiments, delineating successive optimization batches.
    Formulations that could not be synthesized or that underwent deliquescence are indicated by black ``$\times$'' markers; for visualization purposes, these samples are assigned $E_d=\SI{0}{\kilo\joule\per\kilogram}$ and $\mathrm{KPI}=\SI{0}{\USD\per\kWh}$.
    (b) Same dataset sorted by ascending $E_d$, excluding samples that failed synthesis or deliquesced.
    (c) Dataset sorted by descending KPI, again omitting failed samples.
    (d) Resulting Pareto front constructed from all successfully synthesized and characterized composites, separating dominated formulations from the best compromises between high $E_d$ and low KPI. Arrows indicate direction of improvement. Partial point transparency was used so that overlapping markers appear darker, making regions with higher formulation density more visible.
    These plots show how the parallel BO workflow progressively steered experimentation toward regions of the design space exhibiting improved performance trade-offs, while also revealing synthesis boundaries and infeasible formulations.
    }
    \label{fig:opt_results}
\end{figure}

\begin{table}[t]
\centering
\caption{
Formulation parameters, material cost, and screening-level performance indicators of the Pareto-optimal composites identified from the BO campaign.
The reported ${E_d}_{\mathrm{screen}}$ and $\mathrm{KPI}_{\mathrm{screen}}$ values were obtained from the initial screening measurements and were used for Pareto-front construction and candidate selection.
}
\label{tab:selected_composites}
\begin{tabular}{l | c c c c | c c c}
    \thead{Sample} & \thead{Salt} & \thead{$s/s_{max}$ \\ {$\left[\SI{}{\kilogram\per\kilogram}\right]$}} & \thead{$w/c$ \\ {$\left[\SI{}{\kilogram\per\kilogram}\right]$}} & \thead{$a/c$ \\ {$\left[\SI{}{\gram\per\kilogram}\right]$}} & \thead{$C_{\mathrm{mat}}$ \\ {$\left[\SI{}{\USD\per\kg}\right]$}} & \thead{${E_d}_{\mathrm{screen}}$ \\ {$\left[\SI{}{\kilo\joule\per\kilogram}\right]$}} & \thead{$\mathrm{KPI}_{\mathrm{screen}}$ \\ {$\left[\SI{}{\USD\per\kWh}\right]$}} \\
    \hline
    CaCl$_2$-S1       & CaCl$_2$         & 0.50 & 1.00 &  0.0 & 0.06 &  91 &  2.4 \\
    CaCl$_2$-S2       & CaCl$_2$         & 0.90 & 1.43 & 30.0 & 0.14 & 139 &  3.7 \\
    Zn(NO$_3$)$_2$-S1 & Zn(NO$_3$)$_2$   & 0.51 & 1.50 &  0.0 & 0.30 & 154 &  7.0 \\
    Zn(NO$_3$)$_2$-S2 & Zn(NO$_3$)$_2$   & 0.65 & 1.40 &  0.0 & 0.34 & 154 &  7.9 \\
    LiCl-S1           & LiCl             & 0.89 & 1.20 &  4.8 & 3.70 & 482 & 27.8 \\
\end{tabular}
\end{table}

The BO framework enabled a systematic exploration of the four-dimensional formulation space, efficiently guiding the search toward cement--salt composites exhibiting improved trade-offs between thermochemical performance and economic feasibility.
Figure~\ref{fig:opt_results} summarizes the outcomes of the full experimental campaign.
Figure~\ref{fig:opt_results}a reports the experiments in chronological order, highlighting both the diversity of the explored space and the emergence of progressively higher-performing formulations as the optimization advanced.

%\red{
A closer inspection of Figure~\ref{fig:opt_results}a shows that, with the exception of KPI, whose best value was already obtained in the first optimization round for CaCl$_2$-S1, the highest $E_d$ values were identified progressively as the campaign advanced, with the overall maximum being reached in the second last optimization round for the LiCl-S1 composite.
Among the three additional Pareto-optimal formulations located between the lowest-KPI and highest-$E_d$ extremes, one was identified in the second optimization round (Zn(NO$_3$)$_2$-S1), whereas the other two (CaCl$_2$-S2 and Zn(NO$_3$)$_2$-S2) emerged only in the final optimization round.

The late appearance of these intermediate Pareto-optimal composites should not be interpreted as the direct outcome of an explicitly multi-objective optimization strategy, since the BO workflow was implemented as two parallel single-objective pipelines targeting $E_d$ and KPI separately.
Rather, it can be regarded as an emergent consequence of the progressive enrichment of the dataset.
As additional experiments were accumulated around the regions associated with high $E_d$ and low KPI, the surrogate models became increasingly informative in the surrounding portions of the formulation space as well.
This improved local resolution likely facilitated the identification of compositions lying in the compromise region between the two objectives, i.e. formulations that do not optimize either metric individually but provide an attractive trade-off between them.

This aspect is particularly relevant for less intuitive chemistries, such as Zn(NO$_3$)$_2$-based composites, which might have been more difficult to identify using conventional experimental design strategies.
Overall, these observations further highlight the value of the BO-guided workflow in materials exploration under limited experimental budgets, where each new measurement can improve the efficiency and quality of the subsequent search.%}

Several compositions (indicated by black ``$\times$'' markers in Figure~\ref{fig:opt_results}a) either could not be synthesized, typically due to excessively concentrated saline solution that interferes with cement setting and hardening reactions~\cite{lea_leas_2004, lavagna_insight_2022}, or underwent deliquescence during testing.
The latter occurred for three cement--LiCl and three cement--MgCl\textsubscript{2} composites.
To incorporate these failure modes into the BO loop in a numerically stable and informative way, two different penalization strategies were adopted.
For deliquescent samples, the measured $E_d$ was penalized by dividing it by a factor of 50, and the corresponding economic KPI was recomputed accordingly, since its definition depends explicitly on $E_d$ (Equation~\ref{eq:kpi}).
This conservative treatment allowed the surrogate models to identify these regions of the compositional space as highly unfavorable, without removing the corresponding data points from the training set.

The synthesis of cement-based composites was successful for all tested parameters combinations except for one cement-MgCl\textsubscript{2} formulation with an extreme $s/s_{max}$ value of 0.9, which failed to harden and therefore could not be considered a valid composite.
For this formulation, the objective values were manually assigned as $E_d=\SI{1}{\kilo\joule\per\kilogram}$ and $\mathrm{KPI}=\SI{200}{\USD\per\kWh}$ to reflect infeasibility.
For clarity, all failed points are displayed in Figure~\ref{fig:opt_results}a with $E_d=\SI{0}{\kilo\joule\per\kilogram}$ and $\mathrm{KPI}=\SI{0}{\USD\per\kWh}$, and they were excluded from the ranked plots in Figure~\ref{fig:opt_results}b--d.

When the dataset is rearranged by increasing $E_d$ (Figure~\ref{fig:opt_results}b), a clear spread in performance becomes visible, spanning more than an order of magnitude.
Only a limited subset of formulations achieved $E_d>\SI{150}{\kilo\joule\per\kilogram}$, indicating that high $E_d$ is confined to specific regions of the design space rather than being broadly distributed.
Sorting the same experiments by descending economic KPI (Figure~\ref{fig:opt_results}c) reveals a complementary trend: while many formulations exhibit realtively favorable cost metrics, only a smaller fraction simultaneously meets stringent KPI and $E_d$ criteria.
This confirms the intrinsically bi-criterion nature of the formulation problem.

The full set of successfully synthesized and characterized composites was then analyzed \textit{a posteriori} through Pareto-front representation, in order to identify the best compromise between high $E_d$ and low KPI (Figure~\ref{fig:opt_results}d).
The resulting Pareto-optimal formulations, reported in Table~\ref{tab:selected_composites} along with the corresponding formulation parameters, material cost, and screening-stage $E_d$ and economic KPI, identify the non-dominated solutions forming the Pareto frontier, whereas the remaining formulations are sub-optimal in at least one of the two metrics. %\commentAM{Alessio: modificata leggeremnte su suggerimetno di Matteo P.}
Notably, the Pareto front contains representatives from different salt families, indicating that the parallel BO pipelines did not collapse onto a single chemical class but instead explored distinct high-performing regions of the combinatorial space.
Its shape also highlights the expected trade-off between energetic performance and cost: the lowest-cost formulations are associated with moderate specific energy values, whereas the highest-$E_d$ composite incurs a substantially larger economic penalty.
Within this landscape, the BO-guided campaign successfully identified formulations occupying the intermediate trade-off region that balance the two criteria and therefore represent promising candidates for further validation and scale-up.

These results confirm that the proposed BO framework was able to navigate a large and constrained formulation space efficiently, progressively enriching the dataset with informative samples, revealing synthesis boundaries, and identifying a set of cement-salt composites with promising trade-offs between thermochemical performance and economic viability.
In particular, as anticipated in Section~\ref{ssec:optimization}, LiCl, CaCl$_2$, and Zn(NO$_3$)$_2$ appear, to the best of our knowledge, to be previously unexplored active salts in cement-based TCES composites, despite the broader literature on cement--salt hydrate systems~\cite{lavagna_cementitious_2020, salustro_thermal_2024, clark_experimental_2021, skevi_reviewing_2025}.

\subsection{Isotherm Analysis}\label{ssec:isotherms}
\begin{figure}
    \centering
    \includegraphics[width=0.6\linewidth]{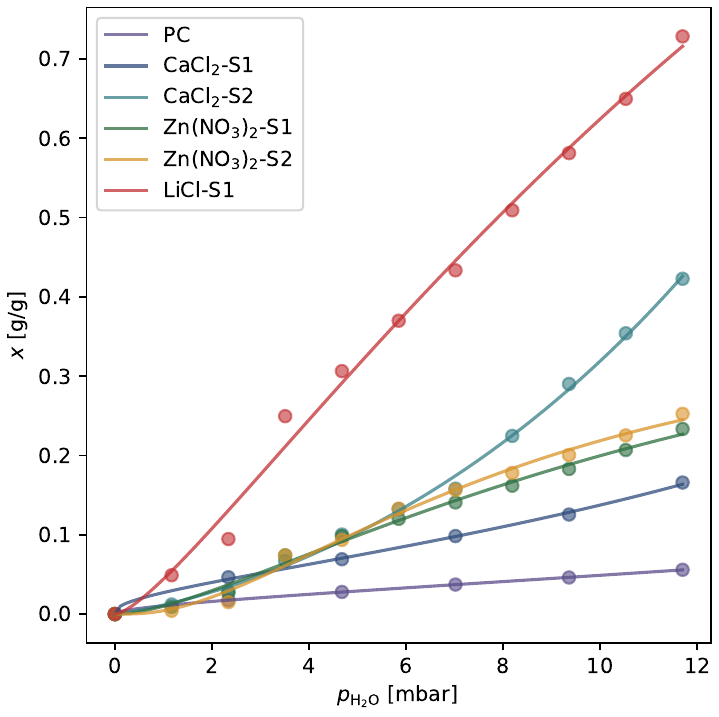}
    \caption{
    Experimental water adsorption isotherms and corresponding Dubinin--Astakhov (DA) fits for the cement reference (PC) and the Pareto-optimal composites identified from the BO campaign.
    Circles denote the experimental DVS data collected at \SI{20}{\degreeCelsius}, while solid lines represent the DA fits.
    The selected composites span different salt families and capture the trade-off between adsorption capacity, specific energy, and economic KPI discussed in the optimization analysis.
    }
    \label{fig:best_results}
\end{figure}

The water adsorption isotherms of the plain cement reference (PC) and of the Pareto-optimal composites are compared in Figure~\ref{fig:best_results}, while the corresponding panel-wise DA fits are reported in the Supplementary Information (Figure~S4).
When the composites are ranked according to water uptake, the best-performing sample is LiCl-S1, which reaches a maximum uptake of 72.8\%, corresponding to \SI{0.73}{\gram\per\gram} of water adsorbed per unit mass of dry composite.
More generally, LiCl-based formulations showed the highest adsorption capacity throughout the screening campaign.
Indeed, among the 20 top-ranked samples in terms of water uptake, fourteen were LiCl-based composites, five were CaCl\textsubscript{2}-based composites, and only one was based on Zn(NO\textsubscript{3})\textsubscript{2}.
The superior performance of LiCl-based composites can be explained by the salt's particularly strong hygroscopic character under the operating conditions considered here, which promotes higher equilibrium water uptake.

The selected composites shown in Figure~\ref{fig:best_results}, however, were not chosen solely on the basis of water uptake.
Rather, they correspond to Pareto-optimal formulations, identified as the most promising trade-offs between high specific energy and low specific energy cost within the BO-guided screening campaign.
As a result, the subset shown in Figure~\ref{fig:best_results} does not necessarily coincide with the top-ranked samples by uptake alone.
For completeness, the individual DA fits for the plain cement reference and for each Pareto-optimal composite are reported separately in the Supplementary Information (Figure~S1).

Nevertheless, all Pareto-optimal composites exhibit a substantially higher uptake at \SI{50}{\percent} RH than the PC matrix, whose uptake is \SI{0.06}{\gram\per\gram}.
The addition of hydrated salts at least doubles the adsorption capacity of the cement matrix and, in the case of LiCl-S1, increases it by approximately one order of magnitude.

Considering the nature of these materials, in which water sorption arises from the combined contributions of salt hydration and adsorption within the cement pore network~\cite{grekova_composite_2017, courbon_further_2017}, the experimental isotherms were fitted using the DA equation (Eq.~\ref{eq:d-a}).
The resulting fitted curves for the Pareto-optimal composites are also shown in Figure~\ref{fig:best_results}.
Overall, a good agreement between the DA fits and the experimental data was observed for all the investigated composites, with $R^2$ exceeding 0.99.%
For the Zn(NO$_3$)$_2$- and LiCl-based samples, a steeper increase in water uptake is observed between the third and fourth experimental points, which is not fully captured by the DA model.
This behavior is likely associated with a change in the hydration state of the salt.
Since chemical hydration phenomena are not explicitly accounted for in the DA formalism, the fitted curves cannot reproduce this feature exactly.
As a consequence, this fitting deviation may lead to a slight underestimation of the cycle uptake variation ($\Delta x_{cycle}$) and therefore of the specific energy.
However, the fitting was considered satisfactory also for these samples, as the deviations between the fitted curves and the experimental data remained limited and did not significantly affect the overall representation of the isotherms.
This is also supported by the corresponding $R^2$ values, which remained higher than 0.99 for these composites.

\begin{figure}[!htbp]
    \centering
    \includegraphics[width=0.70\linewidth]{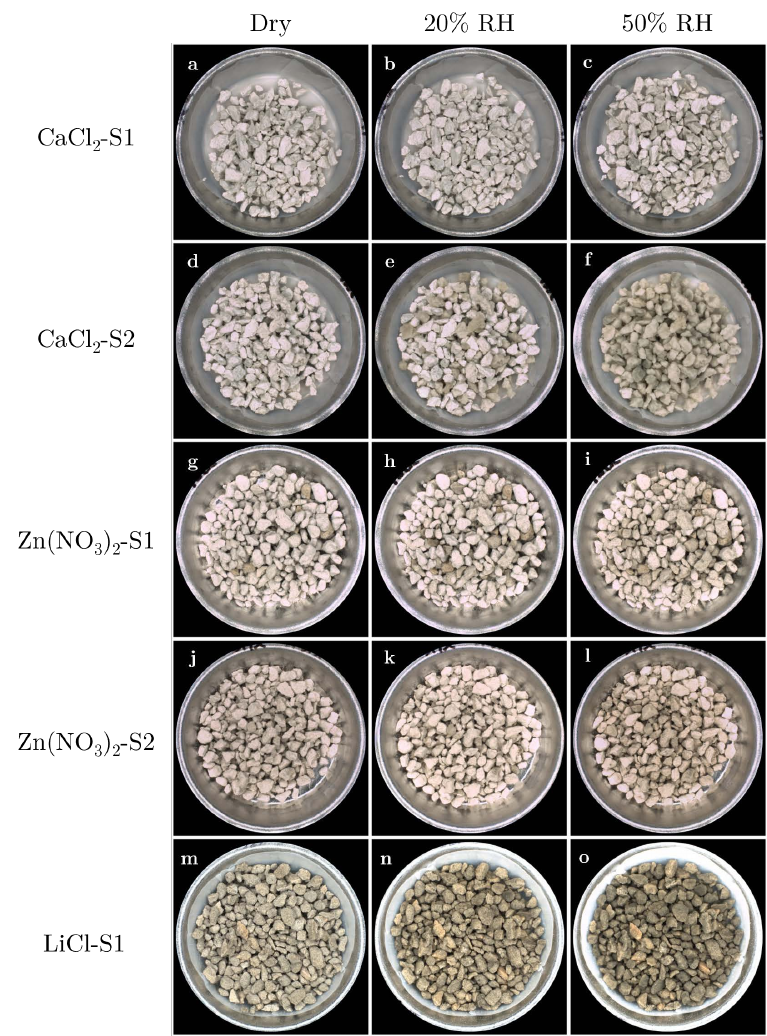}
    \caption{
    Visual appearance of the selected cement-based composites under dry conditions and after equilibration at different relative humidity levels at \SI{20}{\degreeCelsius}.
    Rows correspond to the composite formulations, while columns correspond to the dry state, \SI{20}{\percent} RH, and \SI{50}{\percent} RH.
    The panel labels identify the individual images within the composite--humidity matrix.}
    \label{fig:composite}
\end{figure}

%\red{
A visual comparison of the granulated Pareto-optimal composites under dry conditions and after equilibration at 20\% and 50\% RH is reported in Figure~\ref{fig:composite}.
The dry images were acquired at the beginning of the measurements and are representative of the composites after drying at \SI{90}{\degreeCelsius}.
The images at each RH level were collected after equilibrium had been reached under the corresponding humidity condition.
Among the investigated humidity levels, \SI{20}{\percent} RH was selected as the nearest upper value to the RH expected during the adsorption step considered in this work, namely about \SI{17}{\percent} RH, whereas \SI{50}{\percent} RH was chosen as representative of exposure to high humidity.

At \SI{20}{\percent} RH and \SI{20}{\degreeCelsius}, no clear evidence of salt deliquescence or liquid leakage outside the porous network of the matrix is observed for any of the composites.
However, with the exception of CaCl$_2$-S1 (b), some granules appear visibly darker in all the other formulations (e, h, k, n).
This color change is consistent with water accumulation within the porous network of the composite, most likely in the form of a confined water-rich saline phase inside the granules \cite{grekova_composite_2017, courbon_further_2017}.

At \SI{50}{\percent} RH and \SI{20}{\degreeCelsius}, darker granules also appear in CaCl$_2$-S1 (c), indicating that this formulation starts to exhibit the same visual response already observed at \SI{20}{\percent} RH for the other composites.
For Zn(NO$_3$)$_2$-S1 (i) and Zn(NO$_3$)$_2$-S2 (l), the larger fraction of darker granules suggests a more extensive accumulation of water within the porous network.
In the case of LiCl-S1 (o), almost all granules appear darker, indicating a more uniform and advanced hydration state, possibly approaching the water accommodation capacity of the matrix.
This behavior is even more pronounced for CaCl$_2$-S2 (f), for which all granules exhibit marked darkening.

The near-saturation of the matrix water accommodation capacity for LiCl-S1 and CaCl$_2$-S2 was further confirmed by additional images collected at \SI{60}{\percent} RH, not shown here.
Under these conditions, visible water accumulation was observed at the bottom of the aluminum pan.
This indicates that, at this RH level, the water uptake of the saline phase exceeded the retention capacity of the porous network, causing part of the liquid phase to migrate outside the composite granules.
Conversely, for the other composites, only a progressive increase in the number of darker granules was observed, without clear evidence of water accumulation at the bottom of the pan.

Overall, the most relevant observation is that, under humidity conditions close to those expected during the adsorption process, no clear deliquescence or liquid leakage outside the composite granules was detected.
This suggests that the salt-containing phase remains confined within the cement matrix during adsorption, thereby limiting issues typically associated with salt leakage, such as loss of active material, reduced cycling stability, and degradation of the composite structure \cite{gordeeva_composites_2012, lavagna_cementitious_2020, spietz_thermochemical_2025}.%}

\begin{table}[t]
    \centering
    \caption{
    Repeatability analysis of the DVS measurements for the cement reference and the Pareto-optimal composites re-tested in triplicate.
    For each relative humidity level, the table reports the mean uptake from replicate measurements, $\overline{\Delta x}_{\mathrm{rep}} \; \left[\SI{}{\gram\of{H_2O}\per\gram\of{adsorbent}}\right]$, the corresponding relative standard deviation (RSD) [\SI{}{\percent}], and the absolute deviation (Dev.) $\left[\SI{}{\gram\of{H_2O}\per\gram\of{adsorbent}}\right]$ between the screening measurement and the replicate average.
    }
    \resizebox{\textwidth}{!}
    {
    \begin{tabular}{c|rrr|rrr|rrr|rrr|rrr}
         & \multicolumn{3}{c|}{CaCl$_2$-S1} & \multicolumn{3}{c|}{CaCl$_2$-S2} & \multicolumn{3}{c|}{Zn(NO$_3$)$_2$-S1} & \multicolumn{3}{c|}{Zn(NO$_3$)$_2$-S2} & \multicolumn{3}{c}{LiCl-S1} \\
         RH [\SI{}{\percent}] & $\overline{\Delta x}_\mathrm{{rep}}$ & RSD & Dev. & $\overline{\Delta x}_\mathrm{{rep}}$ & RSD & Dev. & $\overline{\Delta x}_\mathrm{{rep}}$ & RSD & Dev. & $\overline{\Delta x}_\mathrm{{rep}}$ & RSD & Dev. & $\overline{\Delta x}_\mathrm{{rep}}$ & RSD & Dev. \\
         \hline
         5  & 0.021 & 0.7\% & --     & 0.015 & 4.8\% & 0.002 & 0.010 & 7.6\% & 0.001 & 0.005 & 11.1\% & 0.001 & 0.061 & 2.6\% & 0.012 \\
         10 & 0.031 & 0.6\% & 0.016 & 0.034 & 5.1\% & 0.005 & 0.027 & 7.4\% & 0.001 & 0.016 & 0.4\%  & 0.001  & 0.094 & 1.6\% & 0.001 \\
         15 & 0.041 & 0.4\% & --     & 0.074 & 4.2\% & 0.007  & 0.079 & 5.0\% & 0.005 & 0.077 & 1.6\%  & 0.003  & 0.244 & 1.9\% & 0.006 \\
         20 & 0.048 & 0.5\% & 0.021 & 0.107 & 7.5\% & 0.007  & 0.101 & 4.1\% & 0.003 & 0.097 & 1.6\%  & 0.003  & 0.299 & 1.8\% & 0.007 \\
         25 & 0.058 & 0.8\% & --     & 0.138 & 4.2\% & 0.006  & 0.122 & 3.9\% & 0.001 & 0.136 & 1.6\%  & 0.003  & 0.358 & 1.7\% & 0.012 \\
         30 & 0.073 & 0.7\% & 0.025 & 0.167 & 3.1\% & 0.009  & 0.142 & 3.8\% & 0.001 & 0.160 & 1.5\%  & 0.003  & 0.420 & 1.6\% & 0.014 \\
         35 & 0.095 & 0.8\% & --     & 0.237 & 3.0\% & 0.012  & 0.164 & 3.8\% & 0.002 & 0.181 & 1.4\%  & 0.003  & 0.494 & 1.5\% & 0.015 \\
         40 & 0.115 & 1.0\% & 0.011  & 0.291 & 2.6\% & 0.001  & 0.186 & 3.7\% & 0.003 & 0.203 & 1.5\%  & 0.002  & 0.574 & 1.2\% & 0.008 \\
         45 & 0.138 & 0.9\% & --     & 0.352 & 2.3\% & 0.002  & 0.209 & 3.6\% & 0.003 & 0.227 & 1.5\%  & 0.002 & 0.646 & 1.3\% & 0.004 \\
         50 & 0.169 & 0.7\% & 0.003  & 0.426 & 2.0\% & 0.003  & 0.237 & 3.6\% & 0.003 & 0.255 & 1.4\%  & 0.002  & 0.722 & 1.3\% & 0.007 \\
    \end{tabular}
    }
    \label{tab:repeatability}
\end{table}
%

%\red{
To assess the repeatability of the DVS measurements, the Pareto-optimal composites, together with the plain cement reference sample, were re-tested in triplicate. It should be noted that the samples were not re-synthesized; instead, three new aliquots taken from the same batches prepared during the screening stage were tested again. Therefore, this analysis was intended to evaluate the repeatability of the measurements and the robustness of the subsequent modeling, rather than the reproducibility of the synthesis procedure.%} \commentAM{Ho sistemato cosi, dovrebbe essere chiaro adesso}
For each formulation, the mean water uptake and the relative standard deviation (RSD), expressed as a percentage, were calculated from the three replicate measurements according to:
\begin{equation}
    \mathrm{RSD} =  \frac{\sigma_\mathrm{rep}}{\overline{\Delta x}_\mathrm{{rep}}} \times 100
\label{eq:RSD}
\end{equation}
where $\sigma_\mathrm{rep}$ is the standard deviation of the replicate measurements and $\overline{\Delta x}_\mathrm{rep}$ is the corresponding mean water uptake.

The results presented in Table~\ref{tab:repeatability} indicate a good repeatability for all the composites.
The RSD is generally below 5\%, with slightly higher variability observed at low RH, where the absolute water uptake is very small.
Under these conditions, even minor absolute differences among replicates can translate into relatively large percentage variations.
Additional variability in this low-uptake regime may arise from small uptake changes during sample handling and weighing prior to the DVS measurement, which are difficult to avoid completely because of both the high hygroscopicity of the composites and the operating procedure of the instrument.
This effect is particularly evident for the Zn(NO$_3$)$_2$-based composites, which exhibit a sharp uptake increase already at low humidity.
A further contribution may stem from the intrinsic heterogeneity of the granulated composites, including local variations in salt distribution, pore structure, and particle size.

The uptake measured during the initial screening experiment was then compared with the average value obtained from the replicate measurements to evaluate the consistency between the screening-stage measurements and the repeated tests.
The deviation between the two values was defined as:
\begin{equation}
    \mathrm{Dev.}= {\left| \Delta x_{\mathrm{screen}} - \overline{\Delta x}_{\mathrm{rep}} \right|}
\label{eq:Dev}
\end{equation}
where $\Delta x_{\mathrm{screen}}$ is the uptake measured during the screening stage.%}

The agreement between the screening measurements and the replicate averages is generally satisfactory.
The largest deviations are observed for CaCl$_2$-S1, which belongs to the first batch of analysis, and for which both the granulation procedure and the measurement protocol were not yet fully optimized.
The other composites exhibit closer consistency.
Larger deviations are again observed for all the composite at low RH, consistently with the slightly larger RSD values observed for the replicate measurements in the same RH range.
Such behavior can be attributed to the very small absolute water uptake in this region, which makes the results more sensitive to minor absolute differences.
In addition, some water may be adsorbed before the reference mass is fully established, due to the unavoidable exposure of the sample to ambient conditions prior to the measurement.
This introduces a small systematic offset in the measured uptake.
Because the extent of this pre-adsorption may vary slightly from one measurement to another, the resulting offset is systematic but not perfectly reproducible, and therefore becomes particularly relevant at low uptake levels.

\subsection{Thermodynamic Cycle}\label{ssec:cycle}
\begin{figure}
    \centering
    \includegraphics[width=\linewidth]{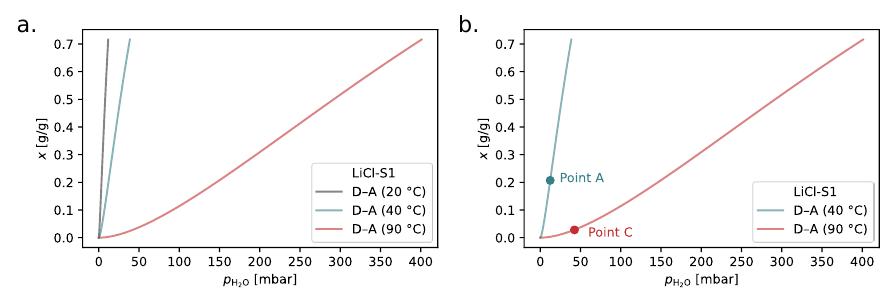}
    \caption{
    Illustration of the thermodynamic-cycle modeling procedure for the LiCl-S1 composite.
    (a) Dubinin--Astakhov (DA) isotherm fitted at \SI{20}{\celsius} and corresponding isotherms shifted to \SI{40}{\celsius} and \SI{90}{\celsius} using Polanyi's potential approach.
    (b) Identification of the two operating points A and C on the shifted isotherms, from which the cycle uptake variation $\Delta x_{\mathrm{cycle}}$ is obtained.
    The same procedure was applied to all composites investigated during the screening.
    }
    \label{fig:combined}
\end{figure}

\begin{table}[t]
\centering
\caption{
Thermodynamic-cycle results and repeatability analysis for the Pareto-optimal composites.
The table reports the cycle uptake variation $\Delta x_{\mathrm{cycle}}$, the screening-level specific energy ${E_d}_{\mathrm{screen}}$, the replicate-based average specific energy $\overline{E_d}_{\mathrm{rep}}$, the corresponding relative standard deviation (RSD), and the replicate-based economic KPI.
}\label{tab:cycle_results}
\begin{tabular}{l | c c c c c}
    \thead{Sample} &
    \thead{{$\Delta x_{\mathrm{cycle}}$} \\ {[\SI{}{\gram\per\gram}]}} &
    \thead{${E_d}_{\mathrm{screen}}$ \\ {[\SI{}{\kilo\joule\per\kilogram}]}} &
    \thead{$\overline{E_d}_{\mathrm{rep}}$ \\ {[\SI{}{\kilo\joule\per\kilogram}]}} &
    \thead{RSD\%} &
    \thead{KPI$_{\mathrm{rep}}$ \\ {[\SI{}{\USD\per\kWh}]}} \\
    \hline
    %PC & 0.014 & 37.5 & 23.4 & 0.4\% & 60.0\% \\%
    CaCl$_2$-S1 & 0.034 & 91 & 69 & 0.6\% & 3.2 \\
    CaCl$_2$-S2 & 0.051 & 139 & 150 & 4.9\% & 3.4 \\
    Zn(NO$_3$)$_2$-S1 & 0.057 & 154 & 158 & 4.3\% & 6.9 \\
    Zn(NO$_3$)$_2$-S2 & 0.057 & 154 & 161 & 1.4\% & 7.6\\
    LiCl-S1 & 0.178 & 482 & 458 & 1.8\% & 29.3 \\
\end{tabular}
\end{table}

From the sorption data, and following the procedure described in Section~\ref{ssec:modelling}, the thermodynamic cycle was evaluated for all composites investigated during the screening campaign.
The resulting screening-level values of $E_d$ and KPI for the Pareto-optimal composites are reported in Table~\ref{tab:selected_composites}.

These values are derived from the initial screening measurements and are primarily used for comparative purposes and candidate selection.
The final and more reliable performance indicators are instead based on the replicate measurements discussed later in this section.
Among the investigated formulations, LiCl-S1 exhibits the highest screening-level ${E_d}_{\mathrm{screen}}$, reaching $\SI{482}{\kilo\joule\per\kilogram}$.
The other Pareto-optimal composites display significantly lower $E_d$ values, nearly fourfold smaller than LiCl-S1.
At the same time, however, these formulations are associated with significantly lower material costs, mainly because LiCl is considerably more expensive than the other salts considered here.
As a result, their lower energy storage performance is at least partly compensated by a more favorable economic KPI.

This trade-off between specific energy and material cost suggests different potential application scenarios.
If maximizing specific energy is the main objective, LiCl-S1 emerges as the most attractive option.
Conversely, for cost-sensitive applications in which lower specific energy can be tolerated and larger adsorber volumes are acceptable, the other Pareto-optimal composites, such as CaCl$_2$-S1/S2 or Zn(NO$_3$)$_2$-S1/S2, may represent more suitable choices.

The transformation from the adsorption isotherms to the thermodynamic-cycle descriptors is illustrated in Figure~\ref{fig:combined} for LiCl-S1, selected here as a representative example.
The same procedure was applied to all screened composites.
Figure~\ref{fig:combined}a shows the DA isotherm fitted at \SI{20}{\celsius}, together with the corresponding isotherms shifted to \SI{40}{\celsius} and \SI{90}{\celsius} according to the procedure described in Section~\ref{ssec:modelling}.
These temperatures correspond to the ambient operating temperature and the charging temperature of the TCES system, respectively.
Figure~\ref{fig:combined}b shows the identification of points A and C on the shifted isotherms, from which the cycle uptake variation $\Delta x_{\mathrm{cycle}}$ is obtained, i.e., the amount of water exchanged during the thermodynamic cycle.
Once the average isosteric heat of adsorption $\overline{q}_{\mathrm{st}}$ had been determined, the cycled heat was calculated using Eq.~\ref{eq:cycled_heat}, allowing the evaluation of $E_d$ and KPI.

The thermodynamic-cycle modeling results for the Pareto-optimal composites are summarized in Table~\ref{tab:cycle_results}.
In addition to the cycle uptake variation, the table reports both the screening-level specific energy and the average value obtained from replicate measurements.
The same repeatability and consistency analysis previously applied to the isotherm data was therefore extended to the derived performance indicator $E_d$.

The results indicate good repeatability for all composites.
The relative standard deviation of $E_d$ remains below \SI{5}{\percent} in all cases, in agreement with the trend already observed for the isotherm measurements.

Likewise, the absolute deviation between the screening-level values of specific energy and the averages obtained from the replicate measurements is generally limited, confirming that the screening procedure provides a reliable basis for comparative ranking despite its high-throughput nature.
The only sample exhibiting a substantially larger deviation is CaCl$_2$-S1, consistently with the behaviour previously observed for its isotherm measurements. In contrast, the other composites show closer agreement between screening-level and replicate-average values.

From this point onward, the performance indicators discussed in the manuscript are based on the replicate measurements reported in Table~\ref{tab:cycle_results}, namely $\overline{E}_{d,\mathrm{rep}}$ and $\mathrm{KPI}_{\mathrm{rep}}$.
This ensures a more robust and statistically reliable assessment of composite performance, while remaining fully consistent with the good agreement observed between screening and replicate values.

\subsection{Performance Comparison}\label{ssec:comparison}
\begin{table}[h]
\centering
\caption{
Comparison between the Pareto-optimal composites identified in this work and representative TCES composites reported in the literature.
For each material, the table reports matrix type (CAC: calcium aluminate cement, Sep.: sepiolite, SG: silica gel, and EV: expanded vermiculite), salt type, salt content, operating conditions adopted for the thermodynamic-cycle evaluation ($p_{ads}$: water vapor pressure during adsorption), estimated material cost, specific energy, and economic KPI.
Only literature data obtained under sufficiently comparable evaluation assumptions were included.}
\label{tab:comparison}
\resizebox{\textwidth}{!}
{
\begin{tabular}{l | c c c c c c c c}
    \thead{Sample} &
    \thead{Matrix} &
    \thead{Salt} &
    \thead{Salt content \\ {[w/w \%]}} &
    \thead{Operating conditions \\ $T_A$ (\SI{}{\celsius}) / $T_C$ (\SI{}{\celsius}) / $p_{ads}$ (\SI{}{\milli\bar})} &
    \thead{$C_{\mathrm{mat}}$ \\ {[\SI{}{\USD\per\kg}]}} &
    \thead{$E_d$ \\ {[\SI{}{\kilo\joule\per\kilogram}]}} &
    \thead{KPI \\ {[\SI{}{\USD\per\kWh}]}} &
    \thead{Ref.}\\
    \hline
    CaCl$_2$-S1 & PC & CaCl$_2$ & 23.0 & 40 / 90 / 12.3 & 0.06 & 69 & 3.2 & This article \\
    CaCl$_2$-S2 & PC & CaCl$_2$ & 39.5 & 40 / 90 / 12.3 & 0.14 & 150 & 3.4 & This article \\
    Zn(NO$_3$)$_2$-S1 & PC & Zn(NO$_3$)$_2$ & 34.4 & 40 / 90 / 12.3 & 0.30 & 158 & 6.9 & This article \\
    Zn(NO$_3$)$_2$-S2 & PC & Zn(NO$_3$)$_2$ & 37.1 & 40 / 90 / 12.3 & 0.34 & 161 & 7.6 & This article \\
    LiCl-S1 & PC & LiCl & 35.8 & 40 / 90 / 12.3 & 3.7 & 458 & 29.3 & This article \\
    PC/MgSO$_4$ & PC & MgSO$_4$ & 21.0 & 35 / 80 / 12.3 & 0.13 - 0.08 & 78 & 3.9 - 6.1 & \cite{lavagna_cementitious_2020} \\
    PCAC-S-1s-B & PC+CAC+Sep. & MgSO$_4$ & 21.5 & 40 / 150 / 12.3 & 0.21 & 84 & 9.3 & \cite{salustro_thermal_2024} \\
    LiCl/Ver & EV & LiCl & 59.0 & 35 / 85 / 12.3 & 6.0 & 2600 & 8.4 & \cite{grekova_composite_2017} \\
    LiCl/Siliaflash & SG & LiCl & 32.4 & 35 / 80 / 8.7 & 18.1 & 1110 & 58.7 & \cite{brancato_experimental_2021} \\
    CaCl$_2$/SG62 & SG & CaCl$_2$ & 43.0 & 40 / 80 / 12.5 & 3.5 & 596 & 21.3 & \cite{courbon_further_2017} \\
\end{tabular}
}
\end{table}

To contextualize the results obtained from the screening procedure, the calculated $E_d$ and KPI values were compared with representative composite materials reported in the literature.
The comparison includes the Pareto-optimal composites identified in this study, two cement-based composites previously investigated by some of the authors of this work (Lavagna et al.~\cite{lavagna_cementitious_2020} and Salustro et al.~\cite{salustro_thermal_2024}), and selected reference composites based on silica gel (SG) and expanded vermiculite (EV).

Since the specific energy considered in this work is evaluated at a system-relevant level, accounting for thermodynamic constraints and operating boundary conditions, only literature data obtained under sufficiently comparable assumptions and methodologies were considered.
This ensures that the comparison remains as consistent and meaningful as possible.
The selected reference materials include two LiCl-based composites, namely LiCl/EV (Grekova et al.~\cite{grekova_composite_2017}) and LiCl/SG (Brancato et al~\cite{brancato_experimental_2021}), as well as a CaCl$_2$/SG composite (Courbon et al.~\cite{courbon_further_2017}).
The corresponding comparison is reported in Table~\ref{tab:comparison}.

A first observation concerns the salt content of the cement-based composites developed in this study, which is generally higher than that reported for previously investigated cement-based systems~\cite{lavagna_cementitious_2020, salustro_thermal_2024, clark_experimental_2021}. 
It should be noted that the salt content reported by Clark et al.~\cite{clark_experimental_2021} was recalculated from the data presented in their article using Eq.~\ref{eq:saltcontent}, in order to ensure consistency with the definition adopted in the present work. This value was found to be 32.6~wt\%, indicating that the composites developed here exhibit only a slightly higher salt content overall, with the exception of the CaCl$_2$-S1 formulation, which shows a lower salt content.
In comparison with the MgSO$_4$-cement composite, the higher salt content can be mainly attributed to the increased $w/c$ ratios enabled by the use of the anti-settling additive, as well as to the higher solubility of CaCl$_2$, LiCl, and Zn(NO$_3$)$_2$ compared to MgSO$_4$.
The salt content of the present materials is broadly comparable to that of SG- and EV-based composites, with the notable exception of the LiCl/Ver sample, which exhibits a remarkably high salt loading of \SI{59}{\percent}.

The operating conditions used for the thermodynamic-cycle evaluation should also be considered carefully when interpreting the comparison.
In particular, the PCAC-S-1s-B composite was evaluated at a charging temperature of \SI{150}{\celsius}, significantly higher than the \SI{90}{\celsius} adopted in this work.
A higher charging temperature generally promotes a larger specific energy \cite{lavagna_cementitious_2020, grekova_composite_2017}, although it also implies a higher energetic (and exergetic) cost during the charging process.
Similarly, most of the literature composites were evaluated at \SI{80}{\celsius}, whereas the materials developed here were evaluated at \SI{90}{\celsius}.
This difference is smaller and is therefore expected to have a more limited effect, but it should nevertheless be kept in mind when comparing absolute $E_d$ values.

A further important aspect concerns the adsorption temperature $T_A$.
Under otherwise identical conditions, a higher adsorption temperature reduces the equilibrium water uptake and therefore decreases the cycle uptake variation and the final cycled heat.
This effect can be substantial.
For instance, Courbon et al.~\cite{courbon_further_2017} reported for a CaCl$_2$/SG composite a reduction in energy storage density from \SI{382}{\kWh\per\meter\cubed} at \SI{20}{\celsius} to \SI{126}{\kWh\per\meter\cubed} at \SI{40}{\celsius}, under otherwise comparable conditions.
In the present work, $T_A$ was fixed at \SI{40}{\celsius}.
As a consequence, compared with literature values obtained at \SI{35}{\celsius}, the composites developed here are expected to be slightly penalized in terms of both $E_d$ and KPI.

The comparison is also instructive when considering the specific salt chemistry.
For the previously reported PC/MgSO$_4$ composite, it should be noted that MgSO$_4$ does not undergo complete dehydration from the heptahydrate to the monohydrate at \SI{80}{\celsius}, but rather remains at the hexahydrate state~\cite{zhong_thermal_2025}.
Therefore, while \SI{80}{\celsius} may be suitable for the other salts considered here, it is not an optimal charging temperature for MgSO$_4$ and may lead to an underestimation of its theoretical thermochemical potential.

Turning to the comparison of the main performance indicators, both the specific energy and the specific energy cost should be considered. 
Regarding the specific energy, it can be observed that the composites developed in this work, supported by the BO, significantly improve the values previously obtained for cement-based composites.
In particular, only the CaCl$_2$-S1 sample exhibits a comparable, although slightly lower, specific energy, while the other composites show values up to twice as high, reaching even an increase by a factor of five in the case of LiCl-S1.
This improvement can be attributed to both the higher salt content achieved and the greater tendency of these cement–salt composites to adsorb water under the investigated conditions compared to cement-MgSO$_4$ composites.
When compared with the best-performing composites reported in the literature~\cite{grekova_composite_2017, courbon_further_2017, brancato_experimental_2021}, the specific energy of the present materials is lower by approximately one order of magnitude for the LiCl-based systems, whereas the difference is less pronounced for the SG62/CaCl$_2$ composite.
These discrepancies can be explained by both the generally higher salt content of literature composites, such as LiCl/Ver, and the significant contribution of the matrix itself to water adsorption, as materials like silica gel or expanded vermiculite are intrinsically more effective adsorbents than cement-based matrices.
It is also worth considering that, since the density of cement is generally higher than that of SG or EV~\cite{lavagna_cementitious_2020, aghemo_comparison_2023}, the volumetric energy density can benefit from this, thereby reducing the differences.

From the point of view of the KPI, the two CaCl$_2$-based composites developed in this study exhibit the lowest values among all the composites considered in this comparison.
The other composites developed in this work, with the exception of LiCl-S1, also show a generally better performance in terms of economic KPI compared to the other materials.
Compared to the previously developed cement-based samples, although the $C_{\mathrm{mat}}$ is comparable, the overall improvement in $E_d$ leads to a more favorable economic KPI.
As for the other literature composites, it is worth highlighting that two of them are strongly penalized by the high cost of LiCl, similarly to the LiCl-S1 composite, while in the case of SG62/CaCl$_2$, the higher $E_d$ does not compensate for the increased cost associated with the SG matrix compared to the cement one.

Overall, the comparison highlights that the cement-based composites developed in this study occupy an intermediate position between previously reported low-cost cementitious materials and high-performance but significantly more expensive SG- or EV-based systems.
In particular, LiCl-S1 provides the highest specific energy among the formulations investigated here, although at a markedly higher material cost, while the CaCl$_2$- and Zn(NO$_3$)$_2$-based composites offer a more balanced compromise between energetic performance and economic viability.
Therefore, these materials represent promising candidates for engineering applications, where the performance-to-cost trade-off is a key aspect.

These results support the potential of BO-guided formulation as a useful strategy for identifying competitive cement-based TCES materials under application-relevant operating conditions.
\section{Conclusions}\label{sec:conclusions}
In this work, we developed a high-throughput BO framework for the formulation of cement--salt hydrate composites for low-temperature TCES applications.
By coupling adaptive experimental design with water-adsorption measurements and thermodynamic-cycle modeling, the proposed workflow enabled the efficient exploration of a four-dimensional formulation space, while also identifying infeasible regions associated with deliquescence or unsuccessful hardening.

The BO-guided campaign led to the identification of Pareto-optimal composites balancing specific energy and economic performance.
In particular, CaCl$_2$-, Zn(NO$_3$)$_2$-, and LiCl-based cement composites emerged as the most promising chemistries within the investigated design space, although they have not been previously explored in literature.
Among them, LiCl-S1 achieved the highest replicate-based specific energy, reaching about $458~\si{kJ.kg^{-1}}$, whereas the CaCl$_2$- and Zn(NO$_3$)$_2$-based formulations exhibited lower but still competitive specific energy values combined with substantially more favorable specific energy costs.
Overall, the optimized materials improved the specific energy of previously developed cement-based composites by up to about a factor of five, confirming that data-driven optimization can accelerate the discovery of higher-performing cementitious TCES materials.

Although the specific energy of the present composites remains below that of state-of-the-art SG- and EV-based systems, several of the formulations identified here offer a more attractive cost-to-performance balance thanks to the low cost of the cement matrix.
The repeatability analysis further confirmed the robustness of the adopted screening workflow and of the derived thermodynamic indicators.
Overall, these results demonstrate both the value of BO as a practical strategy for navigating complex formulation spaces and the potential of cement--salt composites as competitive, low-cost materials for seasonal TCES.
Future work should focus on cycling stability, sorption kinetics, and further formulation refinement, including alternative additives, mixed-salt systems, and matrix modifications aimed at improving accessible porosity and salt dispersion.

\section*{Additional Information}
Supporting Information is available with this article and includes details of the Bayesian optimization workflow, including the 20 BO trial branches adopted at each optimization round; a step-by-step derivation of the thermodynamic-cycle specific energy model based on the Dubinin--Astakhov equation, Polanyi's potential theory, and the Clausius--Clapeyron relation; individual DA fits for the plain cement reference and the Pareto-optimal composites; and a partial experimental validation of the isotherm extrapolation procedure at \SI{40}{\degreeCelsius}.

Portions of the wording in this work were refined with the assistance of ChatGPT, an AI language model by OpenAI, in accordance with the CC-BY 4.0 license; the underlying content was developed by the authors.

\section*{Acknowledgements}
The authors acknowledge financial support from: 
\begin{itemize}
    \item PNRR - M4C2 - AVVISO 341/2022 - NEST - PE00000021 CUP - E13C22001890001 SPOKE 6 - Energy Storage.
    \item EU via Next Generation EU, M4C2, investment 1.1 (project: PRIN PNRR 2022 ``LObSTER'', N. P2022EERT9).
\end{itemize}

\section*{CRediT authorship contribution statement}\label{sec:author-contributions-statement}
\textbf{AM}: Methodology, Software, Formal Analysis, Investigation, Data Curation, Writing -- original draft;
\textbf{GB}: Methodology, Software, Formal Analysis, Visualization, Writing -- original draft;
\textbf{LL}: Methodology, Conceptualization, Supervision, Writing -- review \& editing;
\textbf{MF}: Writing -- review \& editing, Funding acquisition, Project administration;
\textbf{MP}: Conceptualization, Resources, Supervision, Project administration, Funding acquisition, Writing -- review \& editing.
\textbf{EC}: Conceptualization, Resources, Supervision, Project administration, Funding acquisition, Writing -- review \& editing.

All authors read and approved the final manuscript.

\section*{Data Availability}
Data supporting this work is available on Zenodo at \url{https://doi.org/10.5281/zenodo.20184063}.

\section*{Code Availability}
The code used to orchestrate the experimental campaign in this study is publicly available on GitHub at \url{https://github.com/giuliobarl/cementbopt}.

\newpage

\clearpage
% --- reset numbering for supplement ---
\renewcommand{\thefigure}{S\arabic{figure}}
\renewcommand{\thetable}{S\arabic{table}}
\renewcommand{\theequation}{S\arabic{equation}}
\renewcommand{\thesection}{S\arabic{section}}
\renewcommand{\thesubsection}{\thesection.\arabic{subsection}}
\setcounter{figure}{0}
\setcounter{table}{0}
\setcounter{equation}{0}
\setcounter{section}{0}

% --- "frontmatter-like" block (safe centering, wraps correctly) ---
\begingroup
\centering
\setlength{\parindent}{0pt}

% optional: slightly looser line breaking for long lines
\emergencystretch=2em

\begin{minipage}{\textwidth}
\centering
{\LARGE Supplementary Information\par}
\vspace{0.5em}
{\Large High-Throughput Bayesian Optimization of Cement-Salt Hydrates Composites for Seasonal Thermochemical Energy Storage\par}
\vspace{0.75em}
{Alessio Mondello$^{a}$, Giulio Barletta$^{a}$, Luca Lavagna$^{b}$, Matteo Fasano$^{a}$, Matteo Pavese$^{b}$, Eliodoro Chiavazzo$^{a,c,*}$\par}
\vspace{0.5em}
{\small
$^{a}$Department of Energy, Politecnico di Torino, Corso Duca degli Abruzzi, 24, Torino, 10129, Italy\\
$^{b}$Department of Applied Science and Technology, Politecnico di Torino, Corso Duca degli Abruzzi, 24, Torino, 10129, Italy\\
$^{c}$INRIM, Istituto Nazionale di Ricerca Metrologica, Strada delle Cacce, 91, Torino, 10135, Italy\\
}
\end{minipage}
\par
\endgroup

\section{Details of the Bayesian Optimization Workflow} \label{sec:si_bopt_details}
At each optimization round, candidate formulations were generated by combining two optimization objectives, two GP kernels, and five acquisition strategies per kernel.
The resulting 20 BO trial branches are summarized in Table~\ref{tab:bo_branches}.
\begin{table}[t]
    \centering
    \caption[Structure of the 20 BO trial branches considered at each optimization round.]
    {Combinations of objective, GP kernel, and acquisition strategy defining the 20 BO trial branches used at each optimization round.
    For the LCB acquisition function, the exploration parameter $\kappa$ is reported explicitly.
    These branches correspond to the candidate-generation scheme adopted in the two parallel BO pipelines for $E_d$ maximization and KPI minimization.}
    \label{tab:bo_branches}
    \begin{tabular}{cccc}
        \hline
        Objective & GP kernel & Acquisition function & $\kappa$ \\
        \hline
        \multirow{10}{*}{$E_d$}
            & \multirow{5}{*}{RBF}
                & PI  & --- \\
            &   & EI  & --- \\
            &   & LCB & 1.0 \\
            &   & LCB & 4.0 \\
            &   & LCB & 10.0 \\
        \cline{2-4}
            & \multirow{5}{*}{Matérn}
                & PI  & --- \\
            &   & EI  & --- \\
            &   & LCB & 1.0 \\
            &   & LCB & 4.0 \\
            &   & LCB & 10.0 \\
        \hline
        \multirow{10}{*}{KPI}
            & \multirow{5}{*}{RBF}
                & PI  & --- \\
            &   & EI  & --- \\
            &   & LCB & 1.0 \\
            &   & LCB & 4.0 \\
            &   & LCB & 10.0 \\
        \cline{2-4}
            & \multirow{5}{*}{Matérn}
                & PI  & --- \\
            &   & EI  & --- \\
            &   & LCB & 1.0 \\
            &   & LCB & 4.0 \\
            &   & LCB & 10.0 \\
        \hline
    \end{tabular}
\end{table}

\clearpage

\section{Derivation of the Thermodynamic-Cycle Specific Energy Modeling} \label{sec:si_thermo_derivation}
This section provides the detailed derivation of the thermodynamic-cycle modeling procedure adopted to estimate the specific energy of the investigated composites.
In particular, it clarifies how the experimental adsorption isotherms fitted with the Dubinin--Astakhov (DA) equation are combined with Polanyi's potential theory and the Clausius--Clapeyron relation to estimate the average isosteric heat of adsorption and the corresponding cycled heat.

First, it is fundamental to recall the assumption already introduced in the main text, namely that the equilibrium water uptake ($x$) can be expressed as a unique function of the Polanyi adsorption potential, independently of temperature~\cite{aristov_applying_2023}:
\begin{equation}
    x = f(A) .
\label{eq:assumption}
\end{equation}

Several functional forms have been proposed in the literature to relate the equilibrium water uptake to the Polanyi adsorption potential~\cite{aristov_applying_2023}.
In the present work, the Dubinin--Astakhov (DA) equation was adopted in the following form:
\begin{equation}
    x = x_0 \, \exp \left[ - \left( \frac{A}{E} \right)^n \right]
\label{eq:d-a_suppi}
\end{equation}
where $x$ is the equilibrium water uptake, defined as the mass of adsorbed water per unit mass of dry sorbent $\left[\SI{}{\kilogram\of{water}\per\kilogram\of{dry\;sorbent}}\right]$, $x_0$ is the maximum uptake capacity of the material, $A$ is the adsorption potential, $E$ is the characteristic adsorption energy of the system, and $n$ is an empirical parameter.

For an adsorbate in the vapor phase, the Polanyi adsorption potential $A$ is defined as:
\begin{equation}
    A = -RT \ln\left(\frac{p}{p_0}\right)
\label{eq:polanyi_suppi}
\end{equation}
where $R$ is the universal gas constant, $p$ is the equilibrium vapor pressure of the adsorbate, and $p_0$ is the corresponding saturation pressure.

Starting from an experimental isotherm at temperature $T_1$, it is possible to extrapolate, point by point, the corresponding isotherm at another temperature, for example $T_2$.
Considering a point on the isotherm at $T_1$, denoted as $P_1(p_1,x)$, and recalling the assumption expressed in Equation~\ref{eq:assumption}, the Polanyi adsorption potential must be the same at equal equilibrium uptake. Therefore, one can write:
\begin{equation}
    A(T_1,p_1) = A(T_2,p_2)
\label{eq:Aequal}
\end{equation}
Substituting Equation~\ref{eq:polanyi_suppi} into Equation~\ref{eq:Aequal} yields:
\begin{equation}
    -RT_1 \ln\left(\frac{p_1}{p_0(T_1)}\right)
    =
    -RT_2 \ln\left(\frac{p_2}{p_0(T_2)}\right)
\label{eq:polanyi_equal}
\end{equation}
which, when solved for $p_2$, gives the new extrapolated point $P_2(p_2,x)$ on the isotherm at $T_2$.
By applying this procedure to all the points of the experimental isotherm at $T_1$, the entire isotherm can be translated isosterically to $T_2$.

At this stage, the newly obtained points can either be fitted again using the DA equation in Equation~\ref{eq:d-a_suppi}, as done in the present work, or directly described using the fitting parameters obtained for the isotherm at $T_1$, together with the new values of $T_2$ and $p_0(T_2)$ in the expression of $A$.

Two characteristic points, corresponding to the equilibrium water uptake after adsorption and desorption, were then identified on the isotherms associated with the operating temperature $T_A$ and the charging temperature $T_C$, respectively.
Once the thermodynamic boundaries of the system are imposed, the adsorption and desorption pressures, denoted as $p_{\mathrm{ads}}$ and $p_{\mathrm{des}}$, are known from the saturation vapor pressure of water at the corresponding external reference temperatures.
The corresponding equilibrium uptake values, $x_A$ and $x_C$, are then obtained by evaluating the DA isotherms at $p_{\mathrm{ads}}$ and $p_{\mathrm{des}}$, respectively.
Accordingly, the total amount of water exchanged during the cycle can be calculated as:
\begin{equation}
    \Delta x_{\mathrm{cycle}} = x_A - x_C
\end{equation}

For the estimation of the mean isosteric heat of adsorption, the Clausius--Clapeyron relation was written for an isosteric transformation as follows~\cite{trezza_minimal_2022}:
\begin{equation}
\left(\frac{\partial \ln p}{\partial (1/T)}\right)_{x}
=
-\frac{q_{st}}{R}
\label{eq:ccdiff}
\end{equation}
where $q_{st}$ is the isosteric heat of adsorption and $R$ is the universal gas constant.
Integrating Equation~\ref{eq:ccdiff} between two equilibrium states at the same uptake, corresponding to temperatures $T_1$ and $T_2$, gives:
\begin{equation}
\ln\left(\frac{p_2}{p_1}\right)
=
\frac{q_{st}}{R}\left(\frac{1}{T_1}-\frac{1}{T_2}\right)
\end{equation}
and therefore:
\begin{equation}
q_{st}
=
R\frac{T_1T_2}{T_2-T_1}\ln\left(\frac{p_2}{p_1}\right)
\label{eq:qst}
\end{equation}

Since $q_{st}$ may vary with the equilibrium water uptake, five uniformly spaced uptake levels were selected between $x_A$ and $x_C$.
For each uptake level, the corresponding equilibrium water vapor pressures at $T_1$ and $T_2$ were obtained by numerically inverting Equation~\ref{eq:d-a_suppi}.
Five corresponding values of $q_{st}$ were then calculated using Equation~\ref{eq:qst}, and their average was taken as the mean isosteric heat of adsorption of each composite within the framework of the present modeling procedure for the estimation of the cycled heat.

Finally, the cycled heat per unit mass of dry sorbent was calculated as:
\begin{equation}
Q_u=
\overline{q}_{st}
\frac{\Delta x_{cycle}}{M_{H_2O}}
\label{eq:Qu}
\end{equation}
where $M_{H_2O}$ is the molar mass of water. $Q_u$ was finally expressed in $\mathrm{kJ\,kg^{-1}}$ of dry sorbent.

\clearpage

\section{Individual DA Fits for the Reference and Pareto-Optimal Composites} \label{sec:si_individual_fits}
\begin{figure}[h]
    \centering
    \includegraphics[width=\linewidth]{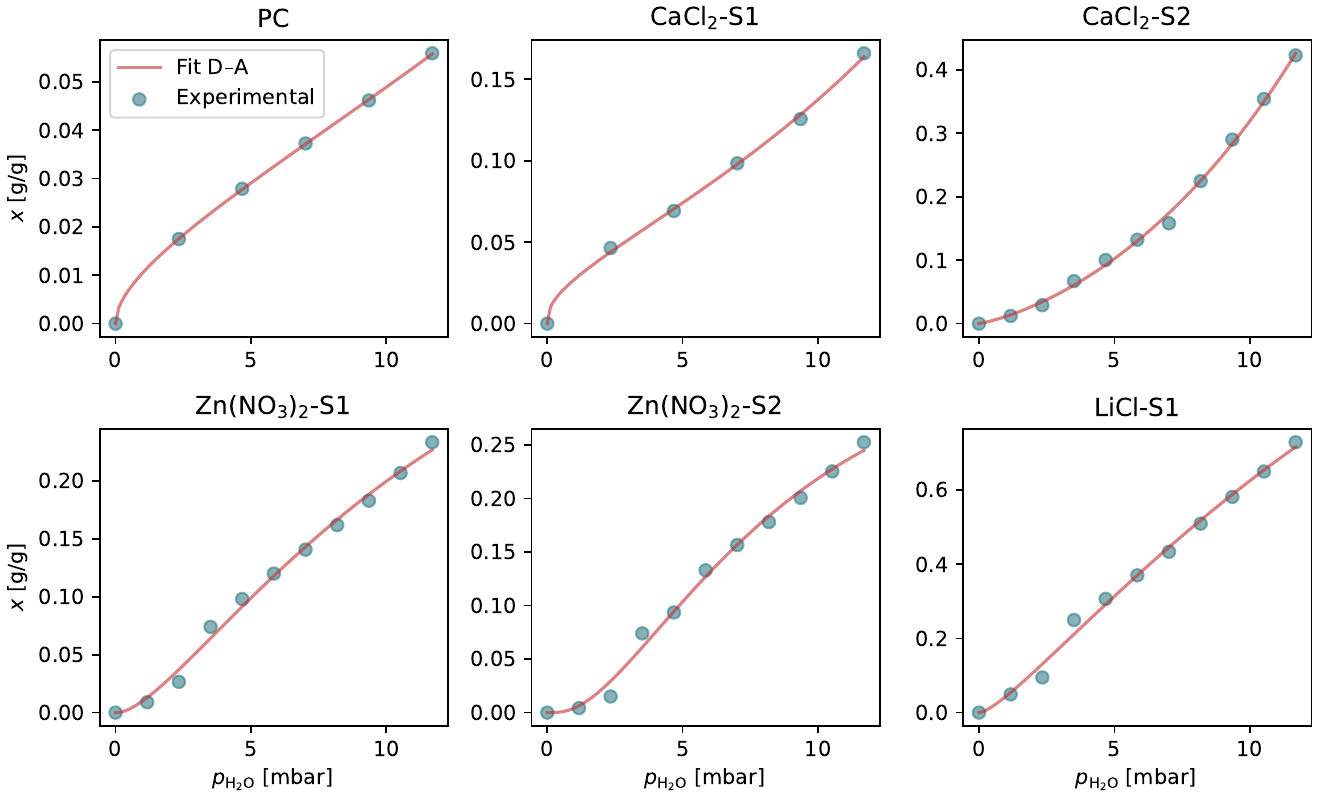}
    \caption{Individual water adsorption isotherms and corresponding Dubinin--Astakhov (DA) fits for the plain cement reference (PC) and the five Pareto-optimal composites identified in the BO campaign.
    Experimental DVS data collected at \SI{20}{\degreeCelsius} are shown as markers, while solid lines represent the DA fits.
    This panel-wise representation complements Figure~4 in the main text, where the same curves are shown together in a single plot to facilitate direct comparison among the selected composites.}
    \label{fig:individual_fits}
\end{figure}

\clearpage

\section{Partial experimental validation of the extrapolated isotherm at \SI{40}{\degreeCelsius}} \label{sec:si_experimental_validation}
In this section of the Supplementary Information, the partial experimental validation of the extrapolated isotherms at \SI{40}{\degreeCelsius} is presented.

The validation is considered partial because it was performed on 11 composite samples only, selected from the initial screening dataset and including all the salts investigated in the study.

The validation was carried out by measuring the water sorption isotherms at \SI{40}{\degreeCelsius} using the same experimental procedure and instrument adopted in the main text for the \SI{20}{\degreeCelsius} isotherms, with the measurement temperature set to \SI{40}{\degreeCelsius}.

\begin{figure}
    \centering
    \includegraphics[width=\linewidth]{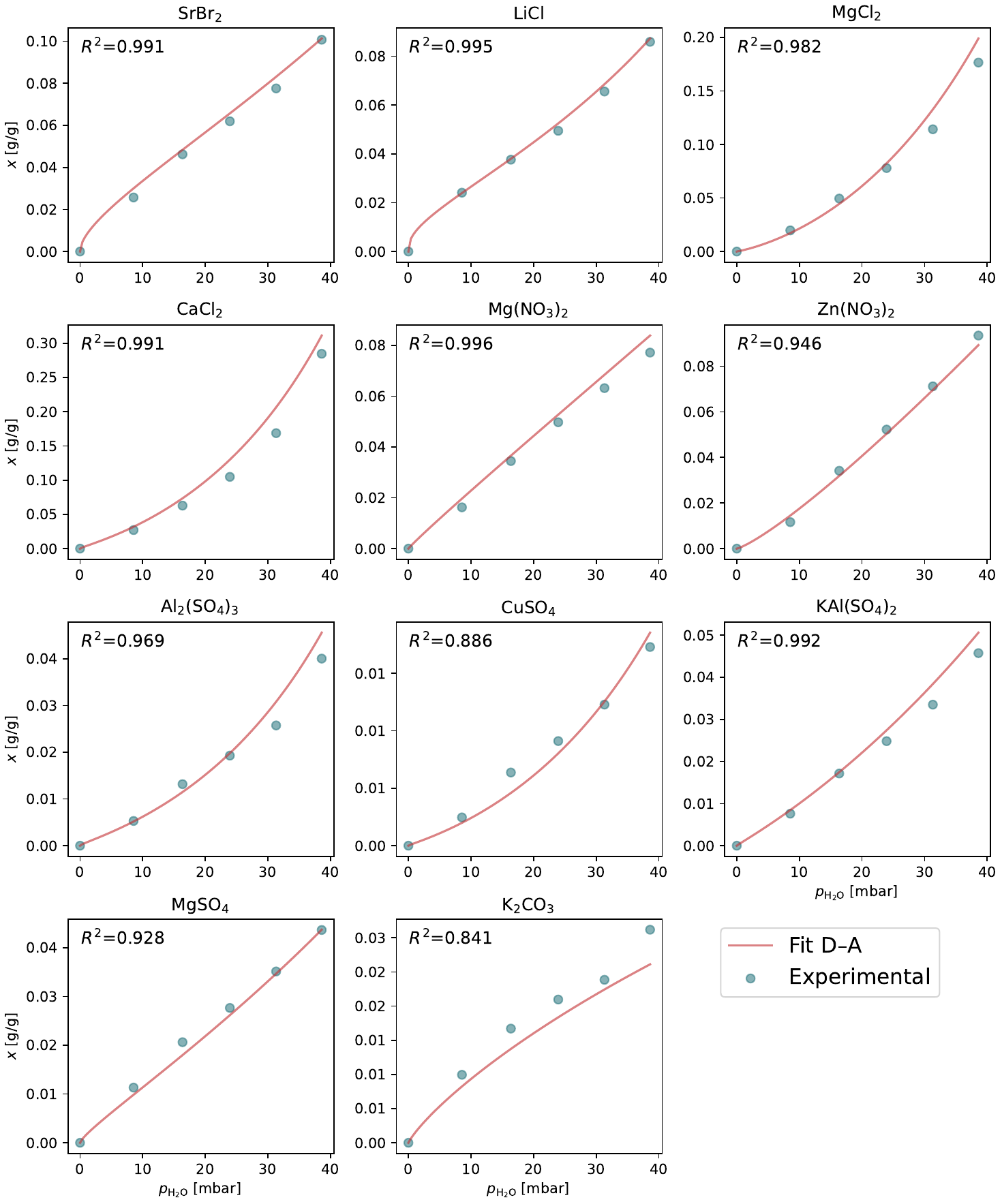}
    \caption{Comparison between the DA isotherms extrapolated to \SI{40}{\degreeCelsius} and the corresponding experimental equilibrium water-uptake data measured at the same temperature for each cement--salt composite.
    Blue markers denote the experimental data, while only the salt name is reported in each subplot for clarity.
    The coefficient of determination, $R^2$, shown in the upper-left corner of each panel provides a quantitative measure of the agreement between the extrapolated isotherm and the corresponding experimental points.}
    \label{fig:validation}
\end{figure}

Figure~\ref{fig:validation} compares, for each composite, the modeled \SI{40}{\degreeCelsius} isotherms obtained from the extrapolation of the \SI{20}{\degreeCelsius} data with the experimental water uptake points measured directly at \SI{40}{\degreeCelsius}.

The $R^2$ value was calculated for each curve as:
\begin{equation}
    R^2 = 1 - \frac{\sum_i \left(x_{i,\mathrm{exp}} - x_{i,\mathrm{mod}}\right)^2}
    {\sum_i \left(x_{i,\mathrm{exp}} - \overline{x}_{\mathrm{exp}}\right)^2}
\label{eq:r2}
\end{equation}
where $x_{i,\mathrm{exp}}$ and $x_{i,\mathrm{mod}}$ are the experimental and modeled water uptake values at the same relative humidity point, respectively, and $\overline{x}_{\mathrm{exp}}$ is the mean experimental uptake.
This coefficient was used to evaluate the agreement between the modeled and experimental isotherms.
The obtained $R^2$ values were satisfactory for most of the investigated composites, indicating that the extrapolated isotherms provide a reasonable description of the experimental behavior at \SI{40}{\degreeCelsius}.

\begin{thebibliography}{10}
\expandafter\ifx\csname url\endcsname\relax
  \def\url#1{\texttt{#1}}\fi
\expandafter\ifx\csname urlprefix\endcsname\relax\def\urlprefix{URL }\fi
\expandafter\ifx\csname href\endcsname\relax
  \def\href#1#2{#2} \def\path#1{#1}\fi

\bibitem{skevi_reviewing_2025}
L.~Skevi, X.~Ke, S.~Ginestet, C.~Ouellet-Plamondon, F.~Gomes, M.~Cyr, \href{https://link.springer.com/10.1617/s11527-025-02803-w}{Reviewing experimental studies on chemical thermal energy storage in {Cementitious} composites: report of the {RILEM} {TC} 299-{TES}}, Materials and Structures 58~(9) (2025) 292.
\newblock \href {https://doi.org/10.1617/s11527-025-02803-w} {\path{doi:10.1617/s11527-025-02803-w}}.

\bibitem{spietz_thermochemical_2025}
T.~Spietz, R.~Fryza, J.~Lasek, J.~Zuwała, \href{https://www.mdpi.com/1996-1073/18/10/2643}{Thermochemical {Energy} {Storage} {Based} on {Salt} {Hydrates}: {A} {Comprehensive} {Review}}, Energies 18~(10) (2025) 2643.
\newblock \href {https://doi.org/10.3390/en18102643} {\path{doi:10.3390/en18102643}}.

\bibitem{whitepaperIWG52024}
W.~van Helden, T.~Cuerdo-Vilches, D.~Lager, A.~Spoden, N.~Cairo, S.~Papa et~al., White paper on clean heating \& cooling technologies and thermal energy storage for buildings (2024).

\bibitem{morciano2025trending}
M.~Morciano, M.~Fasano, E.~Chiavazzo, L.~Mongibello, Trending applications of phase change materials in sustainable thermal engineering: An up-to-date review, Energy Conversion and Management: X 25 (2025) 100862.

\bibitem{neri2020numerical}
M.~Neri, E.~Chiavazzo, L.~Mongibello, Numerical simulation and validation of commercial hot water tanks integrated with phase change material-based storage units, Journal of Energy Storage 32 (2020) 101938.

\bibitem{clark_experimental_2021}
R.-J. Clark, M.~Farid, \href{https://linkinghub.elsevier.com/retrieve/pii/S2352152X21001456}{Experimental investigation into the performance of novel {SrCl2}-based composite material for thermochemical energy storage}, Journal of Energy Storage 36 (2021) 102390.
\newblock \href {https://doi.org/10.1016/j.est.2021.102390} {\path{doi:10.1016/j.est.2021.102390}}.

\bibitem{zhang_research_2025}
X.~Zhang, H.~Xun, Y.~Zhou, Q.~Zhang, R.~Li, X.~Wu et~al., \href{https://linkinghub.elsevier.com/retrieve/pii/S2352152X25017530}{Research progress on thermochemical adsorption heat storage technology of porous matrix loaded hydrated salt}, Journal of Energy Storage 128 (2025) 117040.
\newblock \href {https://doi.org/10.1016/j.est.2025.117040} {\path{doi:10.1016/j.est.2025.117040}}.

\bibitem{liu_systematic_2025}
X.~Liu, F.~Yang, X.~Liu, Y.~Wu, \href{https://linkinghub.elsevier.com/retrieve/pii/S0960148125004938}{A systematic evaluation of sorption-based thermochemical energy storage for building applications: {Material} development, reactor design, and system integration}, Renewable Energy 245 (2025) 122831.
\newblock \href {https://doi.org/10.1016/j.renene.2025.122831} {\path{doi:10.1016/j.renene.2025.122831}}.

\bibitem{yang_salt_2023}
H.~Yang, C.~Wang, L.~Tong, S.~Yin, L.~Wang, Y.~Ding, \href{https://www.mdpi.com/1996-1073/16/6/2875}{Salt {Hydrate} {Adsorption} {Material}-{Based} {Thermochemical} {Energy} {Storage} for {Space} {Heating} {Application}: {A} {Review}}, Energies 16~(6) (2023) 2875.
\newblock \href {https://doi.org/10.3390/en16062875} {\path{doi:10.3390/en16062875}}.

\bibitem{gordeeva_composites_2012}
L.~G. Gordeeva, Y.~I. Aristov, \href{https://academic.oup.com/ijlct/article-lookup/doi/10.1093/ijlct/cts050}{Composites ‘salt inside porous matrix’ for adsorption heat transformation: a current state-of-the-art and new trends}, International Journal of Low-Carbon Technologies 7~(4) (2012) 288--302.
\newblock \href {https://doi.org/10.1093/ijlct/cts050} {\path{doi:10.1093/ijlct/cts050}}.

\bibitem{salustro_thermal_2024}
S.~Salustro, L.~Lavagna, V.~Fernicola, D.~Smorgon, A.~Mondello, E.~Chiavazzo et~al., \href{https://linkinghub.elsevier.com/retrieve/pii/S2352152X24018942}{Thermal characterization and cost analysis of cement-based composite materials for thermochemical energy storage}, Journal of Energy Storage 93 (2024) 112308.
\newblock \href {https://doi.org/10.1016/j.est.2024.112308} {\path{doi:10.1016/j.est.2024.112308}}.

\bibitem{mohapatra_salt_2023}
D.~Mohapatra, J.~Nandanavanam, \href{https://linkinghub.elsevier.com/retrieve/pii/S2214785322037737}{Salt in matrix for thermochemical energy storage - {A} review}, Materials Today: Proceedings 72 (2023) 27--33.
\newblock \href {https://doi.org/10.1016/j.matpr.2022.05.453} {\path{doi:10.1016/j.matpr.2022.05.453}}.

\bibitem{lavagna_insight_2022}
L.~Lavagna, R.~Nisticò, \href{https://www.mdpi.com/2076-3417/13/1/203}{An {Insight} into the {Chemistry} of {Cement}—{A} {Review}}, Applied Sciences 13~(1) (2022) 203.
\newblock \href {https://doi.org/10.3390/app13010203} {\path{doi:10.3390/app13010203}}.

\bibitem{ling_high-dimensional_2017}
J.~Ling, M.~Hutchinson, E.~Antono, S.~Paradiso, B.~Meredig, \href{https://doi.org/10.1007/s40192-017-0098-z}{High-{Dimensional} {Materials} and {Process} {Optimization} {Using} {Data}-{Driven} {Experimental} {Design} with {Well}-{Calibrated} {Uncertainty} {Estimates}}, Integr Mater Manuf Innov 6~(3) (2017) 207--217.
\newblock \href {https://doi.org/10.1007/s40192-017-0098-z} {\path{doi:10.1007/s40192-017-0098-z}}.

\bibitem{lookman_active_2019}
T.~Lookman, P.~V. Balachandran, D.~Xue, R.~Yuan, \href{https://www.nature.com/articles/s41524-019-0153-8}{Active learning in materials science with emphasis on adaptive sampling using uncertainties for targeted design}, npj Comput Mater 5~(1) (2019) 21.
\newblock \href {https://doi.org/10.1038/s41524-019-0153-8} {\path{doi:10.1038/s41524-019-0153-8}}.

\bibitem{gomez-bombarelli_automatic_2018}
R.~Gómez-Bombarelli, J.~N. Wei, D.~Duvenaud, J.~M. Hernández-Lobato, B.~Sánchez-Lengeling, D.~Sheberla et~al., \href{https://doi.org/10.1021/acscentsci.7b00572}{Automatic {Chemical} {Design} {Using} a {Data}-{Driven} {Continuous} {Representation} of {Molecules}}, ACS Cent. Sci. 4~(2) (2018) 268--276.
\newblock \href {https://doi.org/10.1021/acscentsci.7b00572} {\path{doi:10.1021/acscentsci.7b00572}}.

\bibitem{janet_accelerating_2018}
J.~P. Janet, L.~Chan, H.~J. Kulik, \href{https://doi.org/10.1021/acs.jpclett.8b00170}{Accelerating {Chemical} {Discovery} with {Machine} {Learning}: {Simulated} {Evolution} of {Spin} {Crossover} {Complexes} with an {Artificial} {Neural} {Network}}, J. Phys. Chem. Lett. 9~(5) (2018) 1064--1071.
\newblock \href {https://doi.org/10.1021/acs.jpclett.8b00170} {\path{doi:10.1021/acs.jpclett.8b00170}}.

\bibitem{jin_machine-learning-assisted_2022}
W.~Jin, T.~A. Atkinson, C.~Doughty, G.~Neupane, N.~Spycher, T.~L. McLing et~al., \href{https://www.sciencedirect.com/science/article/pii/S0960148122011405}{Machine-learning-assisted high-temperature reservoir thermal energy storage optimization}, Renewable Energy 197 (2022) 384--397.
\newblock \href {https://doi.org/10.1016/j.renene.2022.07.118} {\path{doi:10.1016/j.renene.2022.07.118}}.

\bibitem{frazier_tutorial_2018}
P.~I. Frazier, \href{http://arxiv.org/abs/1807.02811}{A {Tutorial} on {Bayesian} {Optimization}}, arXiv:1807.02811 [stat] (Jul. 2018).
\newblock \href {https://doi.org/10.48550/arXiv.1807.02811} {\path{doi:10.48550/arXiv.1807.02811}}.

\bibitem{brochu_tutorial_2010}
E.~Brochu, V.~M. Cora, N.~d. Freitas, \href{http://arxiv.org/abs/1012.2599}{A {Tutorial} on {Bayesian} {Optimization} of {Expensive} {Cost} {Functions}, with {Application} to {Active} {User} {Modeling} and {Hierarchical} {Reinforcement} {Learning}}, arXiv:1012.2599 [cs] (Dec. 2010).
\newblock \href {https://doi.org/10.48550/arXiv.1012.2599} {\path{doi:10.48550/arXiv.1012.2599}}.

\bibitem{shahriari_taking_2016}
B.~Shahriari, K.~Swersky, Z.~Wang, R.~P. Adams, N.~de~Freitas, \href{https://ieeexplore.ieee.org/abstract/document/7352306}{Taking the {Human} {Out} of the {Loop}: {A} {Review} of {Bayesian} {Optimization}}, Proceedings of the IEEE 104~(1) (2016) 148--175.
\newblock \href {https://doi.org/10.1109/JPROC.2015.2494218} {\path{doi:10.1109/JPROC.2015.2494218}}.

\bibitem{guo_bayesian_2023}
J.~Guo, B.~Ranković, P.~Schwaller, \href{https://www.chimia.ch/chimia/article/view/2023_31}{Bayesian {Optimization} for {Chemical} {Reactions}}, CHIMIA 77~(1-2) (2023) 31--38.
\newblock \href {https://doi.org/10.2533/chimia.2023.31} {\path{doi:10.2533/chimia.2023.31}}.

\bibitem{yang_bayesian_2022}
L.~Yang, A.~Gil, P.~S.~H. Leong, J.~O. Khor, B.~Akhmetov, W.~L. Tan et~al., \href{https://www.sciencedirect.com/science/article/pii/S2352152X22008040}{Bayesian optimization for effective thermal conductivity measurement of thermal energy storage: {An} experimental and numerical approach}, Journal of Energy Storage 52 (2022) 104795.
\newblock \href {https://doi.org/10.1016/j.est.2022.104795} {\path{doi:10.1016/j.est.2022.104795}}.

\bibitem{khatamsaz_bayesian_2023}
D.~Khatamsaz, B.~Vela, P.~Singh, D.~D. Johnson, D.~Allaire, R.~Arróyave, \href{https://www.nature.com/articles/s41524-023-01006-7}{Bayesian optimization with active learning of design constraints using an entropy-based approach}, npj Comput Mater 9~(1) (2023) 49.
\newblock \href {https://doi.org/10.1038/s41524-023-01006-7} {\path{doi:10.1038/s41524-023-01006-7}}.

\bibitem{langner_beyond_2020}
S.~Langner, F.~Häse, J.~D. Perea, T.~Stubhan, J.~Hauch, L.~M. Roch et~al., \href{https://onlinelibrary.wiley.com/doi/abs/10.1002/adma.201907801}{Beyond {Ternary} {OPV}: {High}-{Throughput} {Experimentation} and {Self}-{Driving} {Laboratories} {Optimize} {Multicomponent} {Systems}}, Advanced Materials 32~(14) (2020) 1907801.
\newblock \href {https://doi.org/10.1002/adma.201907801} {\path{doi:10.1002/adma.201907801}}.

\bibitem{agarwal_discovery_2021}
G.~Agarwal, H.~A. Doan, L.~A. Robertson, L.~Zhang, R.~S. Assary, \href{https://doi.org/10.1021/acs.chemmater.1c02040}{Discovery of {Energy} {Storage} {Molecular} {Materials} {Using} {Quantum} {Chemistry}-{Guided} {Multiobjective} {Bayesian} {Optimization}}, Chem. Mater. 33~(20) (2021) 8133--8144.
\newblock \href {https://doi.org/10.1021/acs.chemmater.1c02040} {\path{doi:10.1021/acs.chemmater.1c02040}}.

\bibitem{zhang_exploring_2023}
Y.~Zhang, X.~Yang, C.~Zhang, Z.~Zhang, A.~Su, Y.-B. She, \href{https://www.mdpi.com/2227-9717/11/9/2614}{Exploring {Bayesian} {Optimization} for {Photocatalytic} {Reduction} of {CO2}}, Processes 11~(9) (2023) 2614.
\newblock \href {https://doi.org/10.3390/pr11092614} {\path{doi:10.3390/pr11092614}}.

\bibitem{bonke_multi-variable_2024}
S.~A. Bonke, G.~Trezza, L.~Bergamasco, H.~Song, S.~Rodríguez-Jiménez, L.~Hammarström et~al., \href{https://doi.org/10.1021/jacs.4c01305}{Multi-{Variable} {Multi}-{Metric} {Optimization} of {Self}-{Assembled} {Photocatalytic} {CO2} {Reduction} {Performance} {Using} {Machine} {Learning} {Algorithms}}, J. Am. Chem. Soc. 146~(22) (2024) 15648--15658.
\newblock \href {https://doi.org/10.1021/jacs.4c01305} {\path{doi:10.1021/jacs.4c01305}}.

\bibitem{zhang_optimizing_2023}
J.~Zhang, B.~Liu, Z.~Liu, J.~Wu, S.~Arnold, H.~Shi et~al., \href{https://onlinelibrary.wiley.com/doi/abs/10.1002/aenm.202302594}{Optimizing {Perovskite} {Thin}-{Film} {Parameter} {Spaces} with {Machine} {Learning}-{Guided} {Robotic} {Platform} for {High}-{Performance} {Perovskite} {Solar} {Cells}}, Advanced Energy Materials 13~(48) (2023) 2302594.
\newblock \href {https://doi.org/10.1002/aenm.202302594} {\path{doi:10.1002/aenm.202302594}}.

\bibitem{doan_quantum_2020}
H.~A. Doan, G.~Agarwal, H.~Qian, M.~J. Counihan, J.~Rodríguez-López, J.~S. Moore et~al., \href{https://doi.org/10.1021/acs.chemmater.0c00768}{Quantum {Chemistry}-{Informed} {Active} {Learning} to {Accelerate} the {Design} and {Discovery} of {Sustainable} {Energy} {Storage} {Materials}}, Chem. Mater. 32~(15) (2020) 6338--6346.
\newblock \href {https://doi.org/10.1021/acs.chemmater.0c00768} {\path{doi:10.1021/acs.chemmater.0c00768}}.

\bibitem{niu_accelerated_2025}
X.~Niu, S.~Li, Z.~Zhang, H.~Duan, R.~Zhang, J.~Li et~al., \href{https://doi.org/10.1021/acscatal.5c00467}{Accelerated {Optimization} of {Compositions} and {Chemical} {Ordering} for {Bimetallic} {Alloy} {Catalysts} {Using} {Bayesian} {Learning}}, ACS Catal. 15~(5) (2025) 4374--4383.
\newblock \href {https://doi.org/10.1021/acscatal.5c00467} {\path{doi:10.1021/acscatal.5c00467}}.

\bibitem{yik_accelerating_2025}
J.~T. Yik, C.~Hvarfner, J.~Sjölund, E.~J. Berg, L.~Zhang, \href{https://www.sciencedirect.com/science/article/pii/S266638642500147X}{Accelerating aqueous electrolyte design with automated full-cell battery experimentation and {Bayesian} optimization}, Cell Reports Physical Science 6~(5) (2025) 102548.
\newblock \href {https://doi.org/10.1016/j.xcrp.2025.102548} {\path{doi:10.1016/j.xcrp.2025.102548}}.

\bibitem{li_computational_2025}
W.~Li, Y.~Huang, Q.~Zhang, Z.~Zhang, \href{https://www.sciencedirect.com/science/article/pii/S2590123025016251}{Computational bioinspired structural design for sustainable and secure high-performance energy storage materials}, Results in Engineering 27 (2025) 105555.
\newblock \href {https://doi.org/10.1016/j.rineng.2025.105555} {\path{doi:10.1016/j.rineng.2025.105555}}.

\bibitem{trezza_minimal_2022}
G.~Trezza, L.~Bergamasco, M.~Fasano, E.~Chiavazzo, \href{https://www.nature.com/articles/s41524-022-00806-7}{Minimal crystallographic descriptors of sorption properties in hypothetical {MOFs} and role in sequential learning optimization}, npj Computational Materials 8~(1) (2022) 123.
\newblock \href {https://doi.org/10.1038/s41524-022-00806-7} {\path{doi:10.1038/s41524-022-00806-7}}.

\bibitem{lavagna_cementitious_2020}
L.~Lavagna, D.~Burlon, R.~Nisticò, V.~Brancato, A.~Frazzica, M.~Pavese et~al., \href{https://www.nature.com/articles/s41598-020-69502-0}{Cementitious composite materials for thermal energy storage applications: a preliminary characterization and theoretical analysis}, Scientific Reports 10~(1) (2020) 12833.
\newblock \href {https://doi.org/10.1038/s41598-020-69502-0} {\path{doi:10.1038/s41598-020-69502-0}}.

\bibitem{haynes_crc_2014}
W.~M. Haynes, {CRC} {Handbook} of {Chemistry} and {Physics}, 95th {Edition}, 95th Edition, CRC Press, Hoboken, 2014, oCLC: 908078665.

\bibitem{lin_applications_2021}
J.~Lin, Q.~Zhao, H.~Huang, H.~Mao, Y.~Liu, Y.~Xiao, \href{https://linkinghub.elsevier.com/retrieve/pii/S0038092X20312159}{Applications of low-temperature thermochemical energy storage systems for salt hydrates based on material classification: {A} review}, Solar Energy 214 (2021) 149--178.
\newblock \href {https://doi.org/10.1016/j.solener.2020.11.055} {\path{doi:10.1016/j.solener.2020.11.055}}.

\bibitem{van_essen_characterization_2009}
V.~M. Van~Essen, J.~Cot~Gores, L.~P.~J. Bleijendaal, H.~A. Zondag, R.~Schuitema, M.~Bakker et~al., \href{https://asmedigitalcollection.asme.org/ES/proceedings/ES2009/48906/825/342533}{Characterization of {Salt} {Hydrates} for {Compact} {Seasonal} {Thermochemical} {Storage}}, in: {ASME} 2009 3rd {International} {Conference} on {Energy} {Sustainability}, {Volume} 2, ASMEDC, San Francisco, California, USA, 2009, pp. 825--830.
\newblock \href {https://doi.org/10.1115/ES2009-90289} {\path{doi:10.1115/ES2009-90289}}.

\bibitem{zbair_survey_2021}
M.~Zbair, S.~Bennici, \href{https://www.mdpi.com/1996-1073/14/11/3105}{Survey {Summary} on {Salts} {Hydrates} and {Composites} {Used} in {Thermochemical} {Sorption} {Heat} {Storage}: {A} {Review}}, Energies 14~(11) (2021) 3105.
\newblock \href {https://doi.org/10.3390/en14113105} {\path{doi:10.3390/en14113105}}.

\bibitem{richter_systematic_2018}
M.~Richter, E.-M. Habermann, E.~Siebecke, M.~Linder, \href{https://linkinghub.elsevier.com/retrieve/pii/S0040603117301557}{A systematic screening of salt hydrates as materials for a thermochemical heat transformer}, Thermochimica Acta 659 (2018) 136--150.
\newblock \href {https://doi.org/10.1016/j.tca.2017.06.011} {\path{doi:10.1016/j.tca.2017.06.011}}.

\bibitem{trausel_review_2014}
F.~Trausel, A.-J. De~Jong, R.~Cuypers, \href{https://linkinghub.elsevier.com/retrieve/pii/S1876610214003154}{A {Review} on the {Properties} of {Salt} {Hydrates} for {Thermochemical} {Storage}}, Energy Procedia 48 (2014) 447--452.
\newblock \href {https://doi.org/10.1016/j.egypro.2014.02.053} {\path{doi:10.1016/j.egypro.2014.02.053}}.

\bibitem{donkers_review_2017}
P.~Donkers, L.~Sögütoglu, H.~Huinink, H.~Fischer, O.~Adan, \href{https://linkinghub.elsevier.com/retrieve/pii/S0306261917304750}{A review of salt hydrates for seasonal heat storage in domestic applications}, Applied Energy 199 (2017) 45--68.
\newblock \href {https://doi.org/10.1016/j.apenergy.2017.04.080} {\path{doi:10.1016/j.apenergy.2017.04.080}}.

\bibitem{ntsoukpoe_systematic_2014}
K.~E. N’Tsoukpoe, T.~Schmidt, H.~U. Rammelberg, B.~A. Watts, W.~K. Ruck, \href{https://linkinghub.elsevier.com/retrieve/pii/S0306261914001974}{A systematic multi-step screening of numerous salt hydrates for low temperature thermochemical energy storage}, Applied Energy 124 (2014) 1--16.
\newblock \href {https://doi.org/10.1016/j.apenergy.2014.02.053} {\path{doi:10.1016/j.apenergy.2014.02.053}}.

\bibitem{clark_experimental_2022}
R.-J. Clark, G.~Gholamibozanjani, J.~Woods, S.~Kaur, A.~Odukomaiya, S.~Al-Hallaj et~al., \href{https://linkinghub.elsevier.com/retrieve/pii/S2352152X2200439X}{Experimental screening of salt hydrates for thermochemical energy storage for building heating application}, Journal of Energy Storage 51 (2022) 104415.
\newblock \href {https://doi.org/10.1016/j.est.2022.104415} {\path{doi:10.1016/j.est.2022.104415}}.

\bibitem{seeger_gaussian_2004}
M.~Seeger, \href{https://www.worldscientific.com/doi/abs/10.1142/S0129065704001899}{Gaussian processes for machine learning}, Int. J. Neur. Syst. 14~(02) (2004) 69--106.
\newblock \href {https://doi.org/10.1142/S0129065704001899} {\path{doi:10.1142/S0129065704001899}}.

\bibitem{kushner_new_1964}
H.~J. Kushner, \href{https://doi.org/10.1115/1.3653121}{A {New} {Method} of {Locating} the {Maximum} {Point} of an {Arbitrary} {Multipeak} {Curve} in the {Presence} of {Noise}}, J. Basic Eng 86~(1) (1964) 97--106.
\newblock \href {https://doi.org/10.1115/1.3653121} {\path{doi:10.1115/1.3653121}}.

\bibitem{torn_global_1989}
A.~Törn, A.~Žilinskas (Eds.), \href{https://link.springer.com/10.1007/3-540-50871-6}{Global {Optimization}}, Vol. 350 of Lecture {Notes} in {Computer} {Science}, Springer, Berlin, Heidelberg, 1989.
\newblock \href {https://doi.org/10.1007/3-540-50871-6} {\path{doi:10.1007/3-540-50871-6}}.

\bibitem{jones_taxonomy_2001}
D.~R. Jones, \href{https://doi.org/10.1023/A:1012771025575}{A {Taxonomy} of {Global} {Optimization} {Methods} {Based} on {Response} {Surfaces}}, Journal of Global Optimization 21~(4) (2001) 345--383.
\newblock \href {https://doi.org/10.1023/A:1012771025575} {\path{doi:10.1023/A:1012771025575}}.

\bibitem{mockus_application_1994}
J.~Mockus, \href{https://doi.org/10.1007/BF01099263}{Application of {Bayesian} approach to numerical methods of global and stochastic optimization}, J Glob Optim 4~(4) (1994) 347--365.
\newblock \href {https://doi.org/10.1007/BF01099263} {\path{doi:10.1007/BF01099263}}.

\bibitem{jones_efficient_1998}
D.~R. Jones, M.~Schonlau, W.~J. Welch, \href{https://doi.org/10.1023/A:1008306431147}{Efficient {Global} {Optimization} of {Expensive} {Black}-{Box} {Functions}}, Journal of Global Optimization 13~(4) (1998) 455--492.
\newblock \href {https://doi.org/10.1023/A:1008306431147} {\path{doi:10.1023/A:1008306431147}}.

\bibitem{cox_statistical_1992}
D.~Cox, S.~John, \href{https://ieeexplore.ieee.org/abstract/document/271617}{A statistical method for global optimization}, in: [{Proceedings}] 1992 {IEEE} {International} {Conference} on {Systems}, {Man}, and {Cybernetics}, 1992, pp. 1241--1246 vol.2.
\newblock \href {https://doi.org/10.1109/ICSMC.1992.271617} {\path{doi:10.1109/ICSMC.1992.271617}}.

\bibitem{martin_autonomous_2025}
T.~B. Martin, D.~R. Sutherland, A.~McDannald, A.~G. Kusne, P.~A. Beaucage, \href{https://doi.org/10.1021/acs.chemmater.5c00860}{Autonomous {Small}-{Angle} {Scattering} for {Accelerated} {Soft} {Material} {Formulation} {Optimization}}, Chem. Mater. 37~(12) (2025) 4272--4281.
\newblock \href {https://doi.org/10.1021/acs.chemmater.5c00860} {\path{doi:10.1021/acs.chemmater.5c00860}}.

\bibitem{head_scikit-optimizescikit-optimize_2021}
T.~Head, M.~Kumar, H.~Nahrstaedt, G.~Louppe, I.~Shcherbatyi, \href{https://zenodo.org/records/5565057}{scikit-optimize/scikit-optimize} (Oct. 2021).
\newblock \href {https://doi.org/10.5281/zenodo.5565057} {\path{doi:10.5281/zenodo.5565057}}.

\bibitem{pedregosa_scikit-learn_2011}
F.~Pedregosa, G.~Varoquaux, A.~Gramfort, V.~Michel, B.~Thirion, O.~Grisel et~al., Scikit-learn: {Machine} {Learning} in {Python}, J. Mach. Learn. Res. 12 (2011) 2825--2830.

\bibitem{harris_array_2020}
C.~R. Harris, K.~J. Millman, S.~J. van~der Walt, R.~Gommers, P.~Virtanen, D.~Cournapeau et~al., \href{https://www.nature.com/articles/s41586-020-2649-2}{Array programming with {NumPy}}, Nature 585~(7825) (2020) 357--362.
\newblock \href {https://doi.org/10.1038/s41586-020-2649-2} {\path{doi:10.1038/s41586-020-2649-2}}.

\bibitem{mckinney_data_2010}
W.~McKinney, \href{https://proceedings.scipy.org/articles/Majora-92bf1922-00a}{Data {Structures} for {Statistical} {Computing} in {Python}}, SciPy 2010 (May 2010).
\newblock \href {https://doi.org/10.25080/Majora-92bf1922-00a} {\path{doi:10.25080/Majora-92bf1922-00a}}.

\bibitem{virtanen_scipy_2020}
P.~Virtanen, R.~Gommers, T.~E. Oliphant, M.~Haberland, T.~Reddy, D.~Cournapeau et~al., \href{https://www.nature.com/articles/s41592-019-0686-2}{{SciPy} 1.0: fundamental algorithms for scientific computing in {Python}}, Nat Methods 17~(3) (2020) 261--272.
\newblock \href {https://doi.org/10.1038/s41592-019-0686-2} {\path{doi:10.1038/s41592-019-0686-2}}.

\bibitem{hunter_matplotlib_2007}
J.~D. Hunter, \href{https://ieeexplore.ieee.org/document/4160265}{Matplotlib: {A} {2D} {Graphics} {Environment}}, Computing in Science \& Engineering 9~(3) (2007) 90--95.
\newblock \href {https://doi.org/10.1109/MCSE.2007.55} {\path{doi:10.1109/MCSE.2007.55}}.

\bibitem{grekova_composite_2017}
A.~D. Grekova, L.~G. Gordeeva, Y.~I. Aristov, \href{https://linkinghub.elsevier.com/retrieve/pii/S1359431117302429}{Composite “{LiCl}/vermiculite” as advanced water sorbent for thermal energy storage}, Applied Thermal Engineering 124 (2017) 1401--1408.
\newblock \href {https://doi.org/10.1016/j.applthermaleng.2017.06.122} {\path{doi:10.1016/j.applthermaleng.2017.06.122}}.

\bibitem{courbon_further_2017}
E.~Courbon, P.~D'Ans, A.~Permyakova, O.~Skrylnyk, N.~Steunou, M.~Degrez et~al., \href{https://linkinghub.elsevier.com/retrieve/pii/S0038092X17307144}{Further improvement of the synthesis of silica gel and {CaCl2} composites: {Enhancement} of energy storage density and stability over cycles for solar heat storage coupled with space heating applications}, Solar Energy 157 (2017) 532--541.
\newblock \href {https://doi.org/10.1016/j.solener.2017.08.034} {\path{doi:10.1016/j.solener.2017.08.034}}.

\bibitem{brancato_experimental_2021}
V.~Brancato, L.~G. Gordeeva, A.~Caprì, A.~D. Grekova, A.~Frazzica, \href{https://www.mdpi.com/2073-4352/11/5/476}{Experimental {Comparison} of {Innovative} {Composite} {Sorbents} for {Space} {Heating} and {Domestic} {Hot} {Water} {Storage}}, Crystals 11~(5) (2021) 476.
\newblock \href {https://doi.org/10.3390/cryst11050476} {\path{doi:10.3390/cryst11050476}}.

\bibitem{lea_leas_2004}
F.~M. Lea, P.~C. Hewlett, Lea's chemistry of cement and concrete, 4th Edition, Elsevier-Butterworth-Heinemann, Amsterdam London Paris, 2004.

\bibitem{aristov_applying_2023}
Y.~I. Aristov, \href{https://link.springer.com/10.1007/s10450-023-00385-z}{Applying the potential theory of adsorption for adsorptive heat transformation}, Adsorption 29~(5-6) (2023) 225--235.
\newblock \href {https://doi.org/10.1007/s10450-023-00385-z} {\path{doi:10.1007/s10450-023-00385-z}}.

\bibitem{zhong_thermal_2025}
Y.~Zhong, J.~Li, H.~Wang, M.~Wang, \href{https://linkinghub.elsevier.com/retrieve/pii/S0254058425002597}{Thermal decomposition mechanism of {MgSO4}·{7H2O}}, Materials Chemistry and Physics 337 (2025) 130613.
\newblock \href {https://doi.org/10.1016/j.matchemphys.2025.130613} {\path{doi:10.1016/j.matchemphys.2025.130613}}.

\bibitem{aghemo_comparison_2023}
L.~Aghemo, L.~Lavagna, E.~Chiavazzo, M.~Pavese, \href{https://linkinghub.elsevier.com/retrieve/pii/S2405829722006407}{Comparison of key performance indicators of sorbent materials for thermal energy storage with an economic focus}, Energy Storage Materials 55 (2023) 130--153.
\newblock \href {https://doi.org/10.1016/j.ensm.2022.11.042} {\path{doi:10.1016/j.ensm.2022.11.042}}.

\end{thebibliography}
\end{document}